\documentclass[pdflatex,12pt]{article}
\usepackage{url}
\usepackage{jheppub} 
\usepackage{fancyhdr}
\usepackage{xcolor}
\usepackage{graphicx}
\usepackage{rotating}
\usepackage{comment}

\newcommand{\p}{\partial}
\newcommand{\ali}[1]{\begin{align} #1 \end{align}}
\newcommand{\mn}{{\mu\nu}}
\newcommand{\ra}{\rightarrow}
\newcommand{\vev}[1]{\langle #1 \rangle}

\numberwithin{equation}{section}
\allowdisplaybreaks  

\title{\boldmath Reconstructing bulk equation of motion using CFT modular Hamiltonians}

\author[a]{Nele Callebaut}
\author[b]{and Gilad Lifschytz}

\affiliation[a]{Institute for Theoretical Physics, University of Cologne, Z\"ulpicher Stra{\ss}e 77, 50937 K\"oln, Germany} 
\affiliation[b]{Department of Physics and 
	Haifa Research Center for Theoretical Physics and Astrophysics, 
	University of Haifa, Haifa 31905, Israel} 

\emailAdd{nele.callebaut@thp.uni-koeln.de}  
\emailAdd{giladl@research.haifa.ac.il} 

\abstract{
	In the framework of bulk reconstruction, we elucidate the relationship between the action of CFT modular Hamiltonians on bulk operators, the possible equation of motion for the bulk operators, and the charge distribution at infinity corresponding to such bulk fields.  
	In particular for scalar fields interacting with gravity or with gauge fields, 
	we show how  CFT considerations of the action of the modular Hamiltonian constrain the possible bulk equation of motion to be consistent with bulk gauge invariance and diffeomorphism invariance. In fact we show 
	that requiring that 
	the action of the modular Hamiltonian on a dressed bulk scalar operator be compatible with some unknown simple equation of motion, 
	fixes, 
	under reasonable assumptions, both the equation of motion and the action of the modular Hamiltonian once the first order $\frac{1}{N}$ terms are known.
}

\begin{document} 

\maketitle
	
\pagebreak

\section{Introduction}

In $AdS/CFT$ \cite{Maldacena:1997re},  reconstruction of bulk operators in terms of CFT data or `bulk reconstruction'  
can proceed through various routes. One can start from a bulk point of view, and solve the bulk equation of motion perturbatively in the radial direction, 
with initial conditions on the boundary of AdS \cite{Kabat:2011rz,Heemskerk:2012mn}. This is more a proof of concept since one does not want to use that much  bulk information, and the resulting expressions need to be interpreted in terms of CFT operators. Once bulk operators are constructed from boundary data in terms of CFT operators, one can compute CFT correlation functions and find that they are well-defined and obey appropriate bulk locality conditions. 

The second path is to start with a CFT point of view. One constructs zeroth order expressions \cite{Hamilton:2005ju,Hamilton:2006az}  from symmetry considerations \cite{Nakayama:2015mva,Goto:2017olq} or more generally from intersecting modular Hamiltonian equations \cite{Kabat:2017mun}\footnote{One can also try to use more sophisticated modular flow tools to reconstruct the bulk operator \cite{Faulkner:2017vdd}.}. The zeroth order expression reproduces correctly the bulk two-point function through a CFT computation. 
One then corrects this expression by demanding that inserting the CFT representation of the bulk operator into higher-point CFT correlation functions gives well-defined results \cite{Kabat:2018pbj,Kabat:2020nvj}.  
This results in some expression for the bulk field, from which one can then compute as an output the equation of motion the bulk field obeys \cite{Kabat:2015swa,Kabat:2016zzr,Kabat:2020nvj}. 

Recently, another path was suggested for bulk reconstruction of scalars in $AdS_{3}$ interacting with gauge or gravity fields \cite{Kaplan1708,Chen:2019hdv} (see also \cite{Guica:2015zpf} and \cite{Lewkowycz:2016ukf,Nakayama:2016xvw} for reconstruction proposals in this context). 
The employed condition on the bulk operators comes from demanding a particular boundary distribution of charges, i.e.~a particular form of the commutator between the boundary conserved current and the bulk operator.  
 
Why do all these seemingly different approaches give answers 
that obey the other conditions as a consequence? In this paper, we will see an answer to this question starting to emerge. Central to our discussion will be the action of the modular Hamiltonian. 

In the CFT approach, the action of the modular Hamiltonian on bulk fields is a very useful tool for the reconstruction program. First, the zeroth order term can be derived from intersecting modular Hamiltonian equations. Secondly, for scalar interaction alone, the action of the modular Hamiltonian constrains the correction terms to be particular smearings of primary scalar operators.

What about interactions that come from local symmetries, can the action of the modular Hamiltonian be useful for the reconstruction problem and moreover elucidate the relationship to the approach of \cite{Kaplan1708,Chen:2019hdv}?  
Further, bulk gauge transformations or coordinate transformations that fall off fast enough near the boundary are supposed to be invisible to the boundary CFT. However, these gauge redundancies constrain the possible equation of motion that a bulk field can obey. The equations of motion (in a fixed gauge) should be obeyed by bulk operators reconstructed from the CFT correlation function, and one would like to understand if there is a simple way to understand why (or how) such a structure arises from the CFT point of view, without constructing explicitly well-defined bulk operators and then computing their equation of motion.  

We show in this paper that in the case of a scalar field interacting with gravity or gauge fields, three sets of information about the bulk field are connected by compatibility constraints. These are 1) the equation of motion for the bulk operator, 2) the modular Hamiltonian action 
on the bulk scalar operator, and 3) the charge distribution at infinity of the scalar operator. 
The compatibility 
conditions they must obey tightly constrain any two of the sets given the third. This explains the relationship between the different bulk reconstruction approaches. For instance,  we will show  that knowledge of how the CFT modular Hamiltonian $H_{mod}$ acts on bulk scalar operators interacting with gauge fields or gravity constrains the possible equations of motion the bulk operators can obey, in a way that is consistent with the original bulk local symmetries. In fact we will demonstrate that one can build up consistent pairs of the $H_{mod}$ action on scalar bulk operators and the equation of motion they obey, order by order in $\frac{1}{N}$, resulting in a good bulk operator. 
In addition, a simple charge distribution at infinity can be related to the requirement of  well-defined CFT correlation functions, as in \cite{Kabat:2018pbj}.

In order to  explain  the various  points of views we find useful,  we start with simple examples and work our way up.
In section \ref{sect:gauge} we start with a CFT computation to leading order in $\frac{1}{N}$ of the action of $H_{mod}$ on a scalar bulk operator interacting in $AdS_{3}$ with one abelian Chern-Simons gauge field, dual to a CFT current $j_{-}$. The result is of the general form 
	\ali{
	\frac{1}{2\pi i}[H_{mod},\phi ] = \xi^M \p_M \phi + \Lambda \, \phi \label{generalform} 
}
which we will encounter for each field $\phi$ that we consider, i.e.~dressed by interactions with gauge fields or gravity. The first term is the one expected for a local bulk operator  \cite{JLMS}, while the second term carries information about the non-locality of the dressed field.
From 
the found $H_{mod}$ action on $\phi$, we show there are two possible actions of the Casimir operator on the bulk operator. Comparing both, one gets a correct bulk equation of motion for the scalar bulk operator. This is then generalized in section \ref{sect:allorders} to all orders in $\frac{1}{N}$. In section \ref{sect:jj}  we use a Jacobi identity 
to generalize the construction of section \ref{sect:allorders} to the case where the scalar operator is coupled to two Chern-Simons gauge fields such that the dual CFT has both a $j_{-}$ and a $j_{+}$ conserved current. We then explain that the procedure we used is equivalent to constraining the equation of motion directly by the action of the modular Hamiltonian and reproduce the same results from this new perspective in section \ref{sect:const}. In section \ref{sect:relat} we explain the relationships between different approaches to construct the bulk operator. 
In section \ref{sect:gengauge} we use the point of view of section \ref{sect:const} to constrain the equation of motion for a scalar operator coupled to a bulk gauge field in higher dimensions. This is done using the $H_{mod}$ action on the bulk operator and on the bulk gauge field, and this to all orders in $\frac{1}{N}$ for both the scalar and the gauge field. 
We then argue in section \ref{sect:alt} that our success of getting any consistent bulk equation of motion relies on having a compatible pair  of $H_{mod}$ action and bulk equation of motion, and that one can use the existence of a simple bulk equation of motion as a guiding principle to build up order by order in $\frac{1}{N}$ both the action of $H_{mod}$ on the bulk operator as well as the bulk equation of motion. This is demonstrated in section \ref{gravsection}  with the example of a scalar field interacting with gravity in $AdS_{3}$. We end with some comments about this approach. Some computations are presented in Appendix A-F.

\section{Scalar field interacting with a gauge field  in $AdS_3$\label{sect:gauge}}

We  start with the reconstruction program for a complex scalar field of charge $\frac{q}{N}$, coupled to a gauge field in $AdS_{3}$, with metric 
\ali{
	ds^2 = \frac{1}{Z^2} (dZ^2 + dx^+ dx^-), \label{AdSmetric}
} 
where the gauge field is described by a single Chern-Simons term in the bulk.
This situation is reconstructed in the CFT by a conserved left (or right) moving current $j_{-}$ and a complex scalar operator ${\cal O}$ of dimension $\Delta=2h$ which has a non trivial 3-point function
\begin{equation}
	<{\cal O}(x_1)\bar{{\cal O}}(x_2)j_{-}(x_3)>=-\frac{iq}{N}\frac{1}{(x_{12}^{-}x_{12}^{+})^{\Delta}}\frac{x_{12}^{-}}{x_{13}^{-} x_{23}^{-}} \, , 
\end{equation}
where $x^{\pm}=x\pm t$ and $x_{ij}=x_{i}-x_{j}$.  Our $i\epsilon$ prescription is $t_{j} \rightarrow t_{j}-i\epsilon_{j}$ and the order of operators in the correlation function from left to right is in order of decreasing $\epsilon_{j}$.
Throughout the paper, unless stated otherwise, we normalize operators such that their two-point function is $O(1)$ and the 3-point function is $O(\frac{1}{N})$ where $N^2 \sim$ central charge of the CFT.

Promoting ${\cal O}$ to a bulk field $\phi$ was done in \cite{Kabat:2020nvj} up to $1/N$ corrections,  with the result that the bulk operator up to that order has the form
\begin{equation}
	\phi=\phi^{(0)}+\frac{1}{N}\phi^{(1)}+\cdots \, . 
\end{equation}
The first order correction $\phi^{(1)}=\sum_{l} a_{l} {\cal A}_{(0,l)}$ is represented as a sum of smeared operators of the type
\begin{equation}
	{\cal A}_{(0,l)}=\int K_{\Delta+2+2l} \  \partial^{l+1}_{+} J_{(0,l)}
\end{equation}
where $J_{(0,l)}$ is a primary operator of dimensions $(l+1+\frac{\Delta}{2}, \frac{\Delta}{2})$ of the form
\begin{equation}
	J_{(0,l)}=\sum_{k=0}^{l}  d_{l k}\partial_{-}^{k}j_{-}\partial_{-}^{l-k}{\cal O}
	\label{j0l}
\end{equation}
with
\begin{equation}
	d_{l k}=\frac{(-1)^{k}}{\Gamma(k+1)\Gamma(l-k+1)\Gamma(k+2)\Gamma(l-k+\Delta)}, 
	\label{dk}
\end{equation}
and $K_{\Delta_{l}}$ is the scalar smearing function for a scalar operator of conformal dimension $\Delta_{l}$ obeying $\nabla^2 K_{\Delta_{l}} =\Delta_{l}(\Delta_{l}-2)$. Here,  $\nabla^2 = \nabla_M \nabla^M$ is the box operator in the AdS metric. 

Since the bulk field is coupled to a gauge field one would expect that the physical operator corresponding to it will not commute, due to Gauss law,  with the boundary current 
$j_{0}$, and will thus not transform under AdS isometries as a local bulk scalar. This is manifested in the CFT construction by the fact that the  CFT operators from which the bulk operator is constructed are not smeared primary scalars. In fact it is not possible to construct a primary scalar from a conserved current $j$, a primary scalar $\mathcal O$ and their derivatives for use in the construction at all \cite{Kabat:2012av}. 

It is useful to know how the bulk scalar operator will transform, in particular it is useful to know how  it will transform under the action of the modular Hamiltonian. 
We take the modular Hamiltonian for a segment of length $2R$ centered around the origin\footnote{Here and anywhere else in the paper the modular Hamiltonian is the total modular Hamiltonian for the separation of space into the region $A$ and its complement $\bar{A}$. Then $H_{mod} = H_{mod,A} - H_{mod, \bar A}$.}, in the vacuum state of the CFT. 
It is defined in terms of the CFT generators ($L_1$, $L_{-1}$ and $L_0$, or the translations $P$ and special conformal transformations $Q$) as \cite{Casini:2011kv,deBoer:2016pqk} 
\begin{equation}
	\frac{1}{2\pi i} H_{mod}=\frac{1}{2R}(Q_{0}-R^2P_{0})=\frac{1}{2R} \left((\bar{L}_1-L_1)-R^2(\bar{L}_{-1}-L_{-1}) \right).   \label{Hmoddef}
\end{equation}

The bulk operator was constructed from CFT considerations in \cite{Kabat:2020nvj}. Given the form of the bulk operator as computed from the CFT, we can compute the action of the CFT modular Hamiltonian on it (see Appendix \ref{AppA} for details): 
\begin{equation}
	\frac{1}{2\pi i} [H_{mod},\phi^{(0)}+\frac{1}{N}\phi^{(1)}]=\xi^{M}\partial_{M} (\phi^{(0)}+\frac{1}{N}\phi^{(1)})+\frac{\beta Z^2}{2RN}j_{-}\phi^{(0)},
	\label{hmodact1}
\end{equation}
with 
\ali{ 
	\beta=-i\frac{q}{k}
} 
and $\xi$ the bulk Killing vector defined in \eqref{bulkKilling} that vanishes on the Ryu-Takayanagi surface \cite{Ryu:2006bv} of the considered region. 
From this expression (or computing straightforwardly) one can read off the action of the conformal generators on the bulk operator to be
\begin{equation*}
	[L_{-1}, (\phi^{(0)}+\frac{1}{N}\phi^{(1)})]  =  \partial_{-}(\phi^{(0)}+\frac{1}{N}\phi^{(1)}), \ \  [\bar{L}_{-1}, (\phi^{(0)}+\frac{1}{N}\phi^{(1)})]=\partial_{+}(\phi^{(0)}+\frac{1}{N}\phi^{(1)}), 
\end{equation*}
\begin{equation*}
	[ L_{1}, (\phi^{(0)}+\frac{1}{N}\phi^{(1)}) ]  =  (Zx^{-}\partial_{Z}+(x^{-}){^2}\partial_{-}-Z^2\partial_{+})(\phi^{(0)}+\frac{1}{N}\phi^{(1)} ),
\end{equation*}
\begin{equation*}
	[\bar{L}_1, (\phi^{(0)}+\frac{1}{N}\phi^{(1)}) ]  =  (Zx^{+}\partial_{Z}+(x^{+}){^2}\partial_{+}-Z^2\partial_{-})(\phi^{(0)}+\frac{1}{N}\phi^{(1)})+\beta Z^2j_{-}\phi^{(0)}.
\end{equation*}
From these, the action of $L_{0}$ and $\bar{L}_{0}$ can be computed
\begin{equation*}
	[L_{0}, (\phi^{(0)}+\frac{1}{N}\phi^{(1)} )]=(\frac{1}{2} Z\partial_{Z}+x^{-}\partial_{-})(\phi^{(0)}+\frac{1}{N}\phi^{(1)} ), 
\end{equation*}
\begin{equation*}
	[\bar{L}_{0}, (\phi^{(0)}+\frac{1}{N}\phi^{(1)} )]=(\frac{1}{2} Z\partial_{Z}+x^{+}\partial_{+})(\phi^{(0)}+\frac{1}{N}\phi^{(1)} ).
\end{equation*}
The previous 
expressions can be put into the expression for the quadratic Casimir 
\begin{equation}
	C_{2}=2(L_{0}^{2}+\bar{L}_{0}^{2}-\frac{1}{2}(L_{-1}L_{1}+L_1 L_{-1}+\bar{L}_{-1}\bar{L}_{1}+\bar{L}_1 \bar{L}_{-1}))=\sum_{i,j} (C^{ij}L_{i} L_{j}+\bar{C}^{ij}\bar{L}_{i} \bar{L}_{j}).  \label{c2def}
\end{equation}
We then define the operation
\begin{equation}
C_{2} \circ \phi=\sum_{i,j} (C^{ij}[L_{i}, [L_{j},\phi]]+\bar{C}^{ij}[\bar{L}_{i}, [\bar{L}_{j},\phi]]) . 
\end{equation}

The Casimir operator then has the properties
\begin{equation}
C_{2} \circ \phi^{(0)}=\nabla^2 \phi^{(0)}
\end{equation}
\begin{equation}
	C_{2} \circ (\phi^{(0)}+\frac{1}{N} \phi^{(1)})=\nabla^2 (\phi^{(0)}+\frac{1}{N} \phi^{(1)}) -\frac{2\beta Z^2}{N}j_{-}\partial_{+}\phi^{(0)} . 
	\label{casbasic1}
\end{equation}

We will now see that there is another expression for the action of $C_{2}$ on the bulk operator.
Every ${\cal A}_{(0,l)}$ obeys 
\begin{equation}
	C_{2} \circ {\cal A}_{(0,l)}=(2(l+1+h)(l+h)+2(h(h-1)){\cal A}_{l}=\frac{1}{2}(\nabla^2 +m_{0}^2){\cal A}_{(0,l)} 
\end{equation}
with $m_0^2 = \Delta(\Delta-2)$ the mass of the free scalar $\phi^{(0)}$. 
Since we can also write
\begin{equation}
	C_{2} \circ \phi^{(0)}=\frac{1}{2}(\nabla^2 +m_{0}^2) \phi^{(0)}, 
\end{equation}
we see that
\begin{equation}
	C_{2} \circ (\phi^{(0)}+\frac{1}{N}\phi^{(1)})=\frac{1}{2}(\nabla^2 +m_{0}^2) (\phi^{(0)}+\frac{1}{N} \phi^{(1)}).
	\label{casalt1}
\end{equation}
We can now compare the two expressions for the action of the Casimir. Since they both are correct, it must be that
\begin{equation}
	(\nabla^2 -m_{0}^2)(\phi^{(0)}+\frac{1}{N} \phi^{(1)})-\frac{4\beta Z^2}{N}j_{-}\partial_{+}\phi^{(0)}=0 . 
	\label{eomphi01}
\end{equation}
This is indeed the correct equation of motion a bulk scalar coupled to a gauge field obeys to this order.

\subsection{All orders \label{sect:allorders}}

Here we generalize the above construction to the situation where the bulk scalar gets corrected to all orders in its interaction with the gauge field, but the gauge field does not get back-reacted  
by the scalar field. 
This will give a bulk scalar operator   which is suitable to be inserted into correlation functions with as many $j_{-}$ as we want, but only one additional $\bar{{\cal O}}$, similar to what was done in \cite{Chen:2019hdv}. The simplest possible generalization of (\ref{hmodact1}) is then\footnote{This might not have worked, see section \ref{sect:alt} for a discussion on this point.}
\begin{equation}
	\frac{1}{2\pi i} [H_{mod},\phi]=\xi^{M}\partial_{M}\phi+\frac{\beta Z^2}{2NR}j_{-}\phi . 
	\label{hmodact2}
\end{equation}
where now 
\begin{equation}
\phi=\phi^{(0)}+\sum_{i=1}^{\infty}\frac{1}{N^{i}}\phi^{(i)}.
\end{equation}

Reading off from this the actions of the conformal generators, and inserting those into the expression for the quadratic Casimir, one gets
\begin{equation}
	C_{2}\circ \phi=\nabla^2 \phi  -\frac{2\beta Z^2}{N}j_{-}\partial_{+}\phi.
	\label{casbasic2}
\end{equation}

What CFT operators can we use to build up the $\phi^{(i)}$? 
If we have only $j_{-}$ (and no $j_{+}$), then the only possible multi-trace primary operators we can build from one  ${\cal O}$, $(n+1)$ currents $j_{-}$, and $l$ derivatives are 
the operators $J_{(n,l)}$, which have conformal dimensions $(h+n+l+1, h)$.
The first descendant scalar operator is then $\partial^{n+l+1}_{+}J_{(n,l)}$. Of course one might think that to construct the bulk operators one can  also use smearing of other scalar descendants such as $(\partial_{-} \partial_{+})^{m} \partial^{n+l+1}_{+}J_{(n,l)}$, but this is not compatible with the action of the modular Hamiltonian (\ref{hmodact2})  on the bulk operator.  This is so since equation (\ref{hmodact2}) indicates that $\phi$ transforms under $L_1$ (but not under $\bar{L}_1$) as if it was a smeared primary scalar. Thus  the $\phi^{(n+1)}$  are built from an infinite sum over $l$ of 

\begin{equation}
	{\cal A}_{(n,l)}=\int K_{\Delta+2+2n+2l} \  \partial^{l+n+1}_{+} J_{(n,l)}.
\end{equation}

One can see that, as before,
\begin{equation}
	C_{2} \circ {\cal A}_{(n,l)}=(2(l+1+n+h)(l+h+n)+2h(h-1)){\cal A}_{(n,l)}=\frac{1}{2}(\nabla^2 +m_{0}^2){\cal A}_{(n,l)}.
	\label{cassimp}
\end{equation}
Since only smeared $\partial^{n+l+1}_{+}J_{(n,l)}$ operators 
can appear in $\phi$, 
 we then have
\begin{equation}
	C_{2}\circ \phi=\frac{1}{2}(\nabla^2 +m_{0}^2) \phi.
	\label{cassimpphi}
\end{equation}
Comparing again the two valid expressions for the action of $C_2$ on $\phi$ we find 
\begin{equation}
	(\nabla^2 -m_{0}^2)\phi-\frac{4\beta Z^2}{N}j_{-}\partial_{+}\phi=0 , 
	\label{eomgf}
\end{equation}
which is the right bulk equation of motion when one solves to all orders in $1/N$ for $\phi$ but keeps the bulk gauge field uncorrected beyond the free limit (in $AdS_{3}$ one has $A_{-}(Z,x)\sim j_{-}(x^{-})$ at that order).

\subsection{Including both $j_{-}$ and $j_{+}$\label{sect:jj}}

In this section we look at the situation where the  CFT includes both a left moving and a right moving conserved current $j_{-}$ and $j_{+}$.
We will consider the  situation where the  dual bulk theory has   two Chern-Simons gauge fields $A$ and $\bar{A}$  in a parity preserving combination
\begin{equation}
\frac{k}{2}\ \int \epsilon^{abc} (A_{a}\partial_{b} A_{c}-\bar{A}_{a}\partial_{b} \bar{A}_{c}),
 \end{equation}
  and a scalar field coupled to them with the same charge.
Note that this case, from the scalar field point of view, is more similar to the higher-dimensional case than the situation with just left or right CFT currents.

The simplest expected transformation of the bulk scalar field is then
\begin{equation}
	\frac{1}{2\pi i} [H_{mod},\phi]=\xi^{M}\partial_{M}\phi+\frac{\beta Z^2}{2NR}j_{-}\phi-\frac{\beta Z^2}{2NR}j_{+}\phi . 
	\label{hmodact2+}
\end{equation}
This transformation, to first order in $\frac{1}{N}$, follows from a CFT computation and  can also (to all orders in $\frac{1}{N}$) be derived from the bulk point of view (as explained in Appendix \ref{AppC}). 
Reading from this the action of the conformal generators, and inserting into $C_2$, we get
\begin{equation}
	C_{2} \circ  \phi=\nabla^2 \phi  -\frac{2\beta Z^2}{N}j_{-}\partial_{+}\phi-\frac{2\beta Z^2}{N}j_{+}\partial_{-}\phi . 
	\label{casbasic2+}
\end{equation}

In this case there are many types of primary operators one can build (involving $j_{-}$, $j_{+}$, ${\cal O}$ and derivatives), and use in constructing the $\phi^{(i)}$.  
There does not seem to be any constraint coming from the action of the modular Hamiltonian that
will give a nice equation like (\ref{cassimp}) for each of the involved smeared operators individually.  
The way to move forward is to notice that (\ref{cassimpphi}) cannot be of any form. It has to obey  a boundary condition
and a Jacobi identity. 
The boundary condition is simple, since as $Z \rightarrow 0$ then  $\phi \rightarrow {\cal O}$ and we know that $C_2 \circ {\cal O}=\Delta(\Delta-2){\cal O}$.
The Jacobi identity involves $H_{mod}$, $C_{2}$ and $\phi$, and since in our case $[H_{mod},C_2]=0$, the Jacobi identity becomes
\begin{equation}
	[C_2,[H_{mod},\phi]]|0\rangle =[H_{mod},[C_{2},\phi]]|0 \rangle 
	\label{ji}
\end{equation}
where we have taken the expression to act on the vacuum state of the CFT. Note that since $L_{i}|0>=\bar{L}_{i}|0>=0$,
\begin{equation}
C_{2} \circ \phi |0>=[C_{2},\phi] |0>,
\end{equation}
and as always  $H_{mod}|0>=0$.
Indeed if we take the expressions (\ref{cassimpphi}) and (\ref{hmodact2}), then they both obey the Jacobi identity (\ref{ji}). 

We will now use the Jacobi identity as our guiding principle. If we have both $j_{+}$ and $j_{-}$ so that the action of the modular Hamiltonian is as in (\ref{hmodact2+}), then the action of the Casimir as in (\ref{cassimpphi}) does not obey the Jacobi identity.  
We know that when we have only $j_{-}$ or only $j_{+}$, things reduce to equation (\ref{cassimpphi}). This suggests the simplest ansatz for the Casimir action we can try is ($a$ and $\rho$ are some constants)   
\begin{equation}
	C_{2} \circ \phi=(a \nabla^2 +(1-a) m_{0}^2) \phi-\rho Z^2 j_{+}j_{-} \phi.
	\label{cassimpphi+}
\end{equation}
Using 
\begin{equation}
	\frac{1}{2\pi i}[H_{mod}, j_{+}]=\frac{1}{2R}(2x^{+}j_{+}+(x^{+}+R)(x^{+} -R)\partial_{+} j_{+}),
\end{equation}
we find that this expression for $C_{2}$ will obey the Jacobi identity (\ref{ji}), only  if $a=\frac{1}{2}$ and $\rho=-2\beta^2$.

Comparing now the expressions (\ref{casbasic2+}) and (\ref{cassimpphi+}) gives an equation of motion
\begin{equation}
	(\nabla^2 -m_{0}^2)\phi=\frac{4\beta Z^2}{N}(j_{-}\partial_{+}+j_{+}\partial_{-})\phi-\frac{4\beta^2 Z^2}{N^2}j_{+}j_{-} \phi
	\label{eom2+}
\end{equation}
which is the correct equation of motion for a bulk scalar operator coupled with two (zeroth order) Chern-Simons gauge fields in $AdS_{3}$, in holographic gauge.

One can look for more general ansatze for $C_2 \circ \phi$ obeying the Jacobi identity. 
For instance one can add to the right hand side of (\ref{cassimpphi+}) any polynomial of the form 
\begin{equation}
	\sum_{k} a_{k} \phi^{k+1} \bar{\phi}^{k}  
\end{equation}
provided $\bar{\phi}$ obeys 
\begin{equation}
	\frac{1}{2\pi i} [H_{mod},\bar{\phi}]=\xi^{M}\partial_{M}\bar{\phi}-\frac{\beta Z^2}{2NR}j_{-}\bar{\phi}+\frac{\beta}{2NR}j_{+}\bar{\phi} \, .
	\label{hmodact2+bar}
\end{equation}
This results in a potential term in the equation of motion for the scalar field.

\subsection{Constraints on the equation of motion\label{sect:const}}

Given some known action of a known operator on a bulk field operator, this can constrain the possible equations of motion that the  bulk operator can obey. For instance let us take the action of the modular Hamiltonian on the bulk field to be just
\begin{equation}
	\frac{1}{2\pi i} [H_{mod},\phi]=\xi^{M}\partial_{M}\phi . 
\end{equation}
Let us also assume that the bulk field obeys the equation 
\begin{equation}
	\nabla^2 \phi =m_0^2 \phi . 
\end{equation}
One then must have 
\begin{equation}
	[H_{mod}, \nabla^2 \phi] =m_0^2 [H_{mod}, \phi].
\end{equation}
 
Since $H_{mod}$ does not depend on the coordinates $(Z,x,t)$ of the bulk field, it can go through the differential operator and we get 
\begin{equation}
	\nabla^2 (\xi^{M} \partial_{M} \phi )=m_0^2 \, \xi^{M} \partial_{M} \phi . 
\end{equation}
 That is, the equation of motion must still hold even under modular flow.
This  will only be correct (without further restrictions on $\phi$) if 
\begin{equation}
	[\nabla^2, \xi^{M} \partial_{M}]=0
	\label{coneqb}
\end{equation}
which is correct in our case since $\xi^{M}$ is a Killing vector for the metric that is used to write $\nabla^2$.

Now one can instead take the action of the modular Hamiltonian to be  equation 
(\ref{hmodact2})
\begin{equation}
	\frac{1}{2\pi i} [H_{mod},\phi]=\xi^{M}\partial_{M}\phi+\frac{\beta Z^2}{2NR}j_{-}\phi.
\end{equation}
This time, the previous argument will not hold for the equation of motion 
\begin{equation}
	\nabla^2 \phi -m_{0}^{2} \phi=0. 
	\label{eomf}
\end{equation} 
That is, commuting both sides of the equation of motion with $H_{mod}$ will not give an equality, unless conditions on $\phi$ which are not compatible with the equation of motion are met. So clearly the transformation under $H_{mod}$ and the above equation of motion cannot both be correct.
However, it is not hard to find a different equation of motion that will be compatible with the action of the modular Hamiltonian, and not surprisingly,  the simplest compatible extension of (\ref{eomf}) is the one we found previously
\begin{equation}
	(\nabla^2 -m_{0}^2)\phi-\frac{4\beta Z^2}{N}j_{-}\partial_{+}\phi=0 . 
	\label{coneqright}
\end{equation}
The compatibility means that 
\begin{equation}
	\frac{1}{2\pi i} [H_{mod}, (\nabla^2 -m_{0}^2)\phi-\frac{4\beta Z^2}{N}j_{-}\partial_{+}\phi]=(\xi^{M}\partial_{M}+\frac{\beta Z^2}{2NR}j_{-}) ((\nabla^2 -m_{0}^2)\phi-\frac{4\beta Z^2}{N}j_{-}\partial_{+}\phi).
	\label{compgauge3}
\end{equation}

This line of reasoning is not fundamentally different 
than before 
(but sometimes simpler computationally). The procedure of identifying an equation of motion through two possible expressions of the action of the quadratic Casimir on bulk operators (compatible with the Jacobi identity with $H_{mod}$), results in an equation of motion that guarantees to obey the constraint coming from the known action of $H_{mod}$, such as (\ref{compgauge3}). This is of course extended to the case with both $j_{-}$ and $j_{+}$, where the action of the modular Hamiltonian is given by (\ref{hmodact2+}),  which results in the equation of motion (\ref{eom2+}).

One can look at other constraints, for instance similar constraints can be found by commuting both sides of the equation of motion with $j_{-}(x^-)$. Using 
\begin{equation}
	[j_{-}(x),\phi(Z,y)]=-\frac{q}{N}\delta(x^{-}-y^{-})\phi(Z,y), \ \ [j_{-}(x),j_{-}(y)]=ik \partial_{x^{-}} \delta (x^{-}-y^{-}) , 
\end{equation}
we then find again that equation  
\eqref{eomf}  
is not compatible with this relationship, but equation (\ref{coneqright}) is compatible with it for $\beta= -i\frac{q}{k}$, as before.
Here the compatibility means
\begin{equation}
	[j_{-}(x^{-}),  (\nabla^2 -m_{0}^2)\phi-\frac{4\beta Z^2}{N}j_{-}\partial_{+}\phi]=-\frac{q}{N} \delta(x^{-}-y^{-}) ((\nabla^2 -m_{0}^2)\phi-\frac{4\beta Z^2}{N}j_{-}\partial_{+}\phi).
\end{equation}

Similarly, if one has both $j_{-}$ and $j_{+}$ with
\begin{equation}
	[j_{+}(x),\phi(Z,y)]=\frac{q}{N}\delta(x^{+}-y^{+})\phi(Z,y), \ \ [j_{+}(x),j_{+}(y)]=-ik \partial_{x^{+}} \delta (x^{+}-y^{+}) , 
\end{equation}
then a simple computation shows that  this is compatible with equation (\ref{eom2+}).

\section{Relationships between approaches \label{sect:relat}}

We would like to connect this line of thought to previous investigations of 
the system under consideration. 
In \cite{Chen:2019hdv}, the bulk operator was constructed from a simple,  physically motivated form of its asymptotic charge distribution, which
took the form (in our convention) 
\begin{equation}
	[j_{-}(x),\phi(Z,y)]=-\frac{q}{N}\delta(x^{-}-y^{-})\phi(Z,y).
	\label{chargecond}
\end{equation}
As shown in Appendix \ref{klkapcomp}, the bulk operator computed in  \cite{Kabat:2020nvj} agrees to first order in $\frac{1}{N}$ with the one computed in \cite{Chen:2019hdv}. Thus the condition \eqref{chargecond} leads to an operator obeying the bulk equation of motion (\ref{eomphi01}). 
The connection with the approach using the modular Hamiltonian comes from looking for consistency of the action of the modular Hamiltonian on the bulk field and the commutator of the boundary current with the bulk field.  
The consistency condition is  the Jacobi identity when applied to the operators $H_{mod}$, $j_{-}$ and $\phi$.

Let us assume we know (as before)
\begin{equation}
	\frac{1}{2\pi i}[H_{mod},\phi]=\xi^{M}\partial_{M}\phi+\frac{\beta Z^2}{2NR}j_{-}\phi.
	\label{hmodchar}
\end{equation}
From the CFT we know the $H_{mod}$ action on the current 
\begin{equation}
	\frac{1}{2\pi i}[H_{mod}, j_{-}]=-\frac{1}{2R}(2x^{-}j_{-}+(x^{-}+R)(x^{-} -R)\partial_{-} j_{-}),
\end{equation}
and the commutator of currents 
\begin{equation}
	[j_{-}(x),j_{-}(y)]=ik\partial_{x^{-}}\delta (x^{-}-y^{-}).
\end{equation}

Equation (\ref{chargecond})  is consistent with equation (\ref{hmodchar}) only if it obeys  
the Jacobi identity
\begin{equation}
	[H_{mod},[j_{-},\phi]]+[[H_{mod},\phi], j_{-}]+[\phi,[H_{mod},j_{-}]]=0.
\end{equation}
A simple computation shows that 
it does,  
provided that $\beta=-i\frac{q}{k}$.
But it 
would not be consistent if we took a different action of $H_{mod}$, or a different $[j_{-},\phi]$ with the same $H_{mod}$ action. There of course can be other consistent pairs of $[H_{mod},\phi]$ and $[j_{-},\phi]$.

Thus the action of $H_{mod}$ on $\phi$ is tied to a particular commutator of the conserved current with $\phi$ (i.e  a particular charge distribution at infinity)  and as we saw in previous section, to a particular equation of motion. This explains why the expression for the bulk operator one gets from (\ref{chargecond}) will be the same as the one obeying  the particular equation of motion we got.

\subsection{Connection to $i\epsilon$ issue}
One view of bulk reconstruction is that one  is looking for operators which obey the intersecting modular equations at leading order in $\frac{1}{N}$ and are corrected by operators such that they will produce well-defined correlators. This procedure has many solutions 
that are connected  
by local field redefinitions \cite{Kabat:2015swa,Kabat:2020nvj}.
Indeed just taking $\phi=\phi^{(0)}$ inside a 3-point function with $\bar{{\cal O}}$ and $j_{-}$ produces an ill-defined correlator in the sense that the CFT $i\epsilon$ prescription does not give a unique answer when $\phi^{(0)}$ is put in the middle of the correlator \cite{Kabat:2018pbj}. 
One can view this as a sign that the  commutator of $\phi^{(0)}$ with $j_{-}$ is an ill-defined operator.  
Indeed equation (\ref{jphi0}) shows that the commutator $[j_{-},\phi^{(0)}]$, written in terms of $\phi^{(0)}$,  contains an infinite number of terms with increasing number of derivatives on a delta function, each multiplied by integrals of $\phi^{(0)}$. While this by itself does not mean that the operator is ill-defined, it certainly gives this possibility, while any finite number of local bulk terms will of course produce a well-defined operator. One can then view the condition 
\begin{equation}
	[j_{-}(x^{-}),(\phi^{(0)}+\frac{1}{N}\phi^{(1)})(Z,y^{-},y^{+})]=-\frac{q}{N}\delta(x^{-}-y^{-})\phi^{(0)}
	\label{bjo1}
\end{equation}
as a condition that may guarantee a well-behaved 3-point function 
\begin{equation}
	\langle (\phi^{(0)}+\frac{1}{N}\phi^{(1)}) \bar{\cal O} j_{-} \rangle,
\end{equation}
given that 
\begin{equation}
	\langle \phi^{(0)} \bar{\cal O} \rangle
\end{equation}
is a well-defined correlator. As remarked above, the condition for a well-defined correlator can be satisfied with many forms of $\phi^{(1)}$. For instance another possibility (beyond that used in Appendix \ref{AppA}) discussed in \cite{Kabat:2020nvj}  is, 
\begin{equation}
\phi^{(1)} \rightarrow  \phi^{(1)} +\frac{i \alpha q}{N k}Z^2 j_{-}\partial_{+}\phi^{(0)} 
\end{equation}
with $\alpha$ any constant.  
This results in a different, but still simple charge distribution at infinity
\begin{equation}
[j_{-}(x^{-}),(\phi^{(0)}+\frac{1}{N}\phi^{(1)})(Z,y^{-},y^{+})]=-\frac{q}{N}\delta(x^{-}-y^{-})\phi^{(0)}-\frac{\alpha q}{N} \partial_{-} \delta(x^{-}-y^{-})\partial_{+} \phi^{(0)}.
\end{equation}
This may explain why the approach demanding well-defined correlators \cite{Kabat:2018pbj,Kabat:2020nvj} seems to agree with the approach demanding simple $[j_{-},\phi^{(0)}]$ \cite{Chen:2019hdv}. Further, as seen in the previous section, a specific $[j_{-},\phi^{(0)}]$ results 
in specific equations of motion and a specific $H_{mod}$ action. This also explains why demanding a well-defined $i\epsilon$ prescription results in a bulk operator solving some simple bulk equation of motion. 

We can generalize the charge distribution (\ref{bjo1}) 
to an expression valid for all orders in $\frac{1}{N}$ corrections to $\phi$ 
\begin{equation}
	[j_{-}(x),\phi(Z,y)]=-\frac{q}{N}\delta(x^{-}-y^{-})\phi(Z,y).
	\label{jphicon}
\end{equation}
Then, because of the $\frac{1}{N}$ coefficient in this relationship, we can argue for a well-defined correlator iteratively. 
If we start with $\phi$ to some order in $\frac{1}{N^{k}}$, the commutator only depends on $\phi$ to order $\frac{1}{N^{k-1}}$. So having a solution to (\ref{jphicon}) to order $\frac{1}{N^{k}}$ indicates that the correlators
\begin{equation}
	\langle \phi \bar{\cal O} j_{-}(x^{-}_{1})\cdots j_{-}(x^{-}_{k}) \rangle 
\end{equation}
to  order $\frac{1}{N^{k}}$ will have a well-defined  $i\epsilon$ prescription (and thus 
$\phi$ that solves  (\ref{jphicon}) to $\frac{1}{N^{k}}$ is a well-defined operator)  given that
\begin{equation}
	\langle \phi \bar{\cal O} j_{-}(x^{-}_{1})\cdots j_{-}(x^{-}_{k-1}) \rangle 
\end{equation}
is a well-defined correlator  with $\phi$ a solution of  (\ref{jphicon})  to order $\frac{1}{N^{k-1}}$.

\section{Scalar field interacting with a gauge field in general dimensions \label{sect:gengauge}}

In this section we deal with the higher-dimensional case and all order corrections in $\frac{1}{N}$ for both $\phi$ and $A_{\mu}$. 
We do this using the ideas in section \ref{sect:const}, that is, through looking for equations of motion which are compatible with an assumed  known action of the modular Hamiltonian on the bulk operators.

The assumed action of the modular Hamiltonian ($\frac{1}{2R}(Q_{0}-R^2 P_{0})$) on the bulk operators is
\begin{equation}
	\frac{1}{2\pi i} [H_{mod}, \phi ] = \xi^{M} \partial_{M} \phi +\frac{iq}{2R N} \lambda_{0} \phi
	\label{hmodactgen}
\end{equation}
where 
\begin{equation}
	\xi^{Z}=\frac{Z t}{R}, \ \xi^{i}=\frac{x^{i} t}{R},  \ \xi^{0}=\frac{1}{2R}(Z^2 + t^2 + x_{i}^{2}-R^2) ,
\end{equation} 
and  $\lambda_{0}$ is the compensating gauge transformation parameter \cite{KL1204} obeying $\partial_{Z} \lambda_{0} =2Z A_{0}$, which we can write as 
\begin{equation}
	\lambda_{0}=2\int_{0}^{Z} dZ' Z' A_{0}(Z',x_{i},t). 
\end{equation}
We  in addition need the action of the modular Hamiltonian on the bulk gauge field operator. This is taken from a computation in \cite{Kabat:2018smf},  which was done there for 
the free gauge field case, but we will assume that it is valid in general\footnote{See section \ref{sect:alt} for a logical explanation.}. It is given by 
\begin{equation}
	\frac{1}{2\pi i} [H_{mod}, A_{\mu}] = \xi^{M} \partial_{M} A_{\mu} + A_{\nu}\partial_{\mu} \xi^{\nu} - \frac{1}{2R } \partial_{\mu} \lambda_{0} . 
\end{equation}

The higher-dimensional analogue of equation (\ref{eom2+}) is
\begin{equation}
	(\nabla^2 -m_{0}^{2}) \phi =-2i\frac{q}{N} Z^2 \eta^{\mu \nu}A_{\mu} \partial_{\nu} \phi + \frac{q^2}{N^2} Z^2 \eta^{\mu \nu}A_{\mu} A_{\nu} \phi . 
\end{equation}
We can now check compatibility with the modular Hamiltonian action. Let us label 
\begin{equation}
	EOM_{0}= (\nabla^2 -m_{0}^{2}) \phi +2 i\frac{q}{N} Z^2 \eta^{\mu \nu}A_{\mu} \partial_{\nu} \phi -\frac{q^2}{N^2} Z^2 \eta^{\mu \nu}A_{\mu} A_{\nu} \phi
\end{equation}
for which we indeed find 
\begin{equation}
	\frac{1}{2\pi i} [H_{mod}, EOM_{0}]=(\xi^{M} \partial_{M} +\frac{iq}{2R N} \lambda_{0}) EOM_{0} \,, 
\end{equation}
but only if 
\begin{equation}
	\nabla^2 \lambda_{0} =0. 
	\label{boxlam0}
\end{equation}
This condition is satisfied for $A_{\mu}$ which obey the free gauge field equation of motion. 
This is the analogous case to the $AdS_{3}$ set-up where the corresponding $\lambda_{0}=Z^2 j_{0}$ also  obeys (\ref{boxlam0}). 
To go beyond the free gauge field limit we would need a term that will cancel a term of the form $\phi \nabla^2 \lambda_{0} $, coming from $[H_{mod}, \nabla^2 \phi]$. It is not hard to see that a possible term is of the form
\begin{equation}
	\phi Z^2 \eta^{\mu \nu}\partial_{\mu} A_{\nu}
\end{equation}
which is zero in the free gauge field limit.
Indeed if we now define 
\begin{equation}
	EOM= (\nabla^2 -m_{0}^{2}) \phi +2 i\frac{q}{N} Z^2 \eta^{\mu \nu}A_{\mu} \partial_{\nu} \phi -\frac{q^2}{N^2} Z^2 \eta^{\mu \nu}A_{\mu} A_{\nu} \phi +i\frac{q}{N} \phi Z^2 \eta^{\mu \nu}\partial_{\mu} A_{\nu}
	\label{eomgauge}
\end{equation}
we find 
\begin{equation}
	\frac{1}{2\pi i} [H_{mod}, EOM]=(\xi^{M} \partial_{M} + \frac{iq}{2R N} \lambda_{0}) \  EOM . 
\end{equation}

The equation of motion $EOM=0$ in (\ref{eomgauge}) is of course the correct bulk gauge invariant equation of motion in $A_{Z}=0$ gauge. Thus the imprint of bulk gauge invariance from the CFT point of view can be thought of as a consequence of the $H_{mod}$ action.

\section{An alternative perspective\label{sect:alt}}

In this section we want to present an alternative perspective on the results  of the previous sections. We started with an expression for $[H_{mod},\phi]$ 
in equation (\ref{hmodact1}), which was a result from a CFT computation  to first order in $\frac{1}{N}$. Using this action of the modular Hamiltonian on the bulk operator, we arrived at  
an equation of motion (\ref{eomphi01}) which is compatible with the action of $H_{mod}$. Here there was no surprise. $\phi^{(0)}+\frac{1}{N}\phi^{(1)}$ did solve a simple equation of motion. 
We then made a reasonable guess in (\ref{hmodact2}) as to what the action of the modular Hamiltonian should be when considering higher order correction  terms in reconstructing $\phi$, while still keeping the gauge field reconstruction at lowest order in $\frac{1}{N}$. 
We then used this to get an equation of motion (\ref{eomgf}). This seemed reasonable but in reality it might  have not worked, in the sense that given the guess for the action of the modular Hamiltonian, there was no  guarantee that there would be a compatible equation of motion. Indeed when starting with the action of the modular Hamiltonian (\ref{hmodact2+}), which is the naive generalization of $[H_{mod},\phi]$ computed from the CFT point of view for the case of having both $j_{-}$ and $j_{+}$, one does not end with the naive generalization of the equation of motion. Instead one finds the equation of motion (\ref{eom2+}) which has an extra term proportional to $j_{+}j_{-}\phi$. 

Conceivably, there might not have been an equation of motion that is compatible with (\ref{hmodact2+}) at all. Similar statements hold of course for the pair (\ref{hmodactgen}) and (\ref{eomgauge}). So when we start with an action of the modular Hamiltonian which is appropriate to first order in $\frac{1}{N}$ and a compatible equation of motion to that order, what is the best way to proceed to higher order statements? 

Here we want to suggest that instead of looking at the equation of motion as an output from a compatibility condition with the action of $H_{mod}$, our guiding principle can be to look for a pair of compatible actions. That is, we start with the action of $H_{mod}$ and the equation of motion compatible with it to leading order in $\frac{1}{N}$ and we then try to generalize both of them to higher order in $\frac{1}{N}$, such that they are compatible.  We will see an example of this in the next section\footnote{This may or may not be a unique procedure.}.

\section{Scalar field interacting with gravity in $AdS_{3}$}  \label{gravsection}

In this section we consider a 2-dimensional CFT with a scalar operator $\mathcal O$ 
that couples to a conserved CFT stress tensor $T_\mn$, corresponding in the dual theory to a gravitationally dressed scalar bulk operator $\phi$ in an  asymptotically $AdS_3$ background.  
That is, the CFT stress tensor is dual to metric perturbations $h_{\mn}$ of the 
background, that the dual scalar $\phi$ interacts with. 

We set out to determine the $H_{mod}$ action on the CFT representation of the gravitationally dressed field $\phi$. The action of $H_{mod}$ captures the non-locality of the physical, i.e.~diffeomorphism invariant $\phi$, and we will show that demanding compatibility of the $H_{mod}$ action with the existence of a 
bulk equation of motion is enough to determine both the $H_{mod}$ action on $\phi$ and the bulk equation of motion for $\phi$ order by order in $1/N$, once the first term linear it $T_{\mu \nu}$ is known.

In the case of a conserved, chiral current of general spin $s$ being present in the CFT, the form of $\phi^{(1)}$ was derived from CFT principles in \cite{Kabat:2020nvj}. Making use of this known expression, the $H_{mod}$ action on $\phi = \phi^{(0)} + \frac{1}{N} \phi^{(1)}$ was determined in Appendix \ref{AppB} to be \eqref{Hmodjs}.    
In the presence of the spin 2 conserved current $T_\mn$ in the CFT, 
the result of \eqref{Hmodjs} 
applies to 
linear order, with a contribution from each chirality of the stress tensor as in equation \eqref{hmodact2+}. That is,
\ali{
	\frac{1}{2\pi i} [H_{mod},(\phi^{(0)} + \frac{1}{N} \phi^{(1)})] = \xi^M \p_M (\phi^{(0)} + \frac{1}{N} \phi^{(1)})+ \frac{Z^4}{2R} \frac{\gamma}{N} \tilde{T}_{--}\p_+ \phi^{(0)} -  \frac{Z^4}{2R} \frac{\gamma}{N} \tilde{T}_{++}\p_- \phi^{(0)} ,   \label{HmodgravCFT}
}
with $\xi$ the Killing vector \eqref{bulkKilling} of $AdS_3$. It is important to note that the normalization leading to this equation is such that all 2-point functions are $O(1)$, without any $N \sim  \sqrt{c}$ dependence, where
$c$ is the CFT central charge. However, the two-point function of the energy momentum tensor $\sim c$ and this is why in the above equation there appears $\tilde{T}_{\pm \pm} =\frac{1}{N} T_{\pm \pm}$. 
The constant $\gamma \sim O(1)$ in \eqref{Hmodjs} was left undetermined in Appendix \ref{AppB}. We will choose normalizations that fix the constant $\frac{\gamma}{N^2} = \frac{6}{c}$ at the end of this section.

This $H_{mod}$ action is of the general 
form \eqref{generalform}, with the first term  
the contribution as if $\phi$ is just a local  bulk field, and the second and third term expressing the non-locality of the gravitationally interacting $\phi$ 
in the form of stress tensor dependent derivative operators $\Lambda(T)$. 
Because of the non-linearity of gravity,  
the $H_{mod}$ action 
can be expected to receive contributions at each order in $1/N$, 
\ali{
	\frac{1}{2\pi i} [H_{mod},\phi] =\sum_{i=0}^\infty \left(\frac{1}{N}\right)^i 
	\mathcal L_H^{(i)} \phi = \mathcal L_H \phi ,  \label{generalHgrav}
}  
having introduced the notation $\mathcal L_H^{(i)}$ for the $i$-th order in $\tilde T_{\pm\pm}$ derivative operator acting on the all-order field $\phi$, and $\mathcal L_H$ for the corresponding total derivative operator. The total field $\phi$ is expanded as $\sum_{i=0}^\infty (\frac{1}{N})^i \phi^{(i)}$. 
To first order, equation \eqref{generalHgrav} is just a rewriting of \eqref{HmodgravCFT} to 
\ali{
	\mathcal L_H^{(0)} = \xi^M \p_M, \qquad \mathcal L_H^{(1)} =   \frac{\gamma}{2R} 
	\left( Z^4 \tilde T_{--} \p_+ - Z^4 \tilde T_{++} \p_- \right) \label{LH0LH1}  
}
or 
\ali{
	\frac{1}{2\pi i} [H_{mod},\phi^{(0)}] = \mathcal L_H^{(0)} \phi^{(0)}, \qquad 
	\frac{1}{2\pi i} [H_{mod},\phi^{(1)}] = \mathcal L_H^{(0)} \phi^{(1)}+\mathcal L_H^{(1)} \phi^{(0)}. \label{firstordersHgrav}
}

The modular Hamiltonian action $[H_{mod},\phi]$ in the CFT knows about - or at least strongly constrains - the bulk equation of motion for $\phi$.  
Namely, as in equation \eqref{compgauge3}, we impose 
that the bulk equation of motion should satisfy the compatibility condition 
\ali{
	\frac{1}{2\pi i} [H_{mod},EOM] =  \mathcal L_H \, EOM .  	\label{Hmodconditiongrav}
}
Now it would have been enough if on the right hand side of (\ref{Hmodconditiongrav}) there would appear any differential operator, but since we expect that there will be a linear term with arbitrary constant coefficient in $\phi$ in the equation of motion, this linear operator must be $ \mathcal L_H$. 
With the knowledge of the $H_{mod}$ action to first order in $1/N$, given in \eqref{HmodgravCFT} or \eqref{LH0LH1}-\eqref{firstordersHgrav}, 
this condition 
indeed allows us to determine the bulk equation of motion to first order in $1/N$. This we will now first demonstrate. 

We take as an ansatz for the unknown bulk equation of motion for the gravitationally interacting field $\phi$, the form 
\ali{
	\nabla^2 \phi = RHS, \qquad RHS = \sum_{i=0}^\infty \left(\frac{1}{N}\right)^i RHS^{(i)}   \label{EOMansatz}  
}
with right hand side terms $RHS$, which we expand in a priori unknown $i$-th order in $\frac{1}{N}$ terms. At zeroth order, the condition \eqref{Hmodconditiongrav} is consistent with  
\ali{ 
	RHS^{(0)} = b \, \phi^{(0)}  \label{RHS0}
}
for any constant $b$, if $\nabla^2$ in \eqref{EOMansatz} is the box operator in $AdS_3$ such that it commutes with $\mathcal L_H^{(0)}$.  

With this ansatz, the condition \eqref{Hmodconditiongrav} becomes 
\ali{	
	\nabla^2 (\mathcal L_H \phi) - \frac{1}{2\pi i}[H_{mod},RHS] = \mathcal L_H \nabla^2 \phi - \mathcal L_H \, RHS.  	\label{conditiongrav}
}

\subsection{First order}

At first order in $1/N$, 
the condition \eqref{conditiongrav} reads  
\ali{
	\mathcal L_H^{(0)} RHS^{(1)} - \frac{1}{2\pi i}[H_{mod},RHS^{(1)}] 
	= \mathcal L_H^{(1)} \nabla^2 \phi - \nabla^2 \mathcal L_H^{(1)} \phi - \mathcal L_H^{(1)} RHS^{(0)} .   \label{eq323}
}
The right hand side of this condition is known and given by (with primes for derivatives)  
\ali{
	\begin{split} \hspace{-0.4cm}	&\frac{2Z^4 \gamma}{R} \left(  Z^2 \tilde T_{++}' \p_-^2  - Z^2 \tilde T_{--}' \p_+^2   + 2 \tilde T_{++} \p_- - 2 \tilde T_{--} \p_+ + 2 Z \tilde T_{++} \p_Z \p_-   - 2 Z \tilde T_{--} \p_Z\p_+  \phantom{\frac{b}{4}} \right.  \\ 
		& \qquad \left. + \frac{b}{4} (\tilde T_{++}\p_- - \tilde T_{--}\p_+) \right)\phi^{(0)}.   \label{rhseq323}
	\end{split} 
}

We can build an ansatz for $RHS^{(1)}$ (and later on for all $RHS^{(n)}, n>0$) based on the following principles: Lorentz invariance in the $AdS_3$ conformal boundary indices $\mn$, which take the values $+$ and $-$ (i.e.~the terms in the ansatz should contain equal numbers of $+$ and $-$ indices),  
symmetry in exchanging $+$ and $-$ indices, and dimensional analysis (to fix the accompanying orders of $Z$). It turns out to be sufficient for finding a solution to include at most second order derivatives on $\phi$ and $\tilde T_{\pm\pm}$ in the ansatze. We have not investigated including higher order derivatives. The ansatz for $RHS^{(1)}$ is then given by $a Z^4 (\tilde T_{++} \p_-^2 \phi^{(0)} + \tilde T_{--} \p_+^2 \phi^{(0)}) + b \, \phi^{(1)}$, with the coefficient $a$ an undetermined constant. The ansatz also includes the expected first order correction $b \, \phi^{(1)}$ to $RHS^{(0)}$ given in \eqref{RHS0}, with $b$ a mass parameter that can take on any value. 

With this ansatz, the left hand side of \eqref{eq323} can be worked out  using 
the action of $H_{mod}$ on $RHS^{(1)}$ given in terms of 
the action on $\phi$ in \eqref{HmodgravCFT} and that on the stress tensors as 
\ali{
	\begin{split} 
		[H_{mod},RHS^{(1)}] &= \p_{\tilde T_{++}} RHS^{(1)} \, [H_{mod},\tilde T_{++}] + \p_{\tilde T_{--}} RHS^{(1)} \, [H_{mod},\tilde T_{--}] +  b  \,  [H_{mod},\phi^{(1)}] \\ 
		&\, + \p_{\p_-^2\phi^{(0)}} RHS^{(1)} \, \p_-^2 [H_{mod},\phi^{(0)}] +  \p_{\p_+^2\phi^{(0)}} RHS^{(1)} \, \p_+^2 [H_{mod},\phi^{(0)}] . 
	\end{split}      
}
The action of $H_{mod}$ on the CFT stress tensors $[H_{mod},\tilde T_{\pm\pm}]$ is immediate from the $H_{mod}$ definition \eqref{Hmoddef} and the action of the conformal generators on the stress tensor, given explicitly in \eqref{L1Tpp}-\eqref{L1Tmm}. 
The left hand side of \eqref{eq323} then takes a form that matches the right hand side \eqref{rhseq323} if and only if $a=4 \gamma$ or 
\ali{
	RHS^{(1)} = 4 Z^4 \gamma \, (\tilde T_{++} \p_-^2 \phi^{(0)} + \tilde T_{--} \p_+^2 \phi^{(0)}) + b \, \phi^{(1)}.  \label{RHS1known}
}

\subsection{Second order}

We only have access at this stage to the CFT obtained modular Hamiltonian action \eqref{HmodgravCFT}, valid to first order in $1/N$. That is, in the notation of equation \eqref{generalHgrav}, the derivative operators $\mathcal L_H^{(i)}$ are unknown for $i \geq 2$. We can proceed by \emph{assuming}  that the modular Hamiltonian action \eqref{HmodgravCFT} is valid also to the next order, i.e.~that $\mathcal L_H^{(2)} = 0$\footnote{This was indeed the case, in the gauge field case.}. (This assumption will turn out to be false.) If we do so, the compatibility condition with the existence of a bulk equation of motion \eqref{Hmodconditiongrav} at second order in $1/N$ takes the form  
\ali{
	\begin{split} \mathcal L_H^{(0)} RHS^{(2)} - \frac{1}{2\pi i}[H_{mod},RHS^{(2)}] &= \mathcal L_H^{(2)} \nabla^2 \phi^{(0)} - \nabla^2 \mathcal L_H^{(2)} \phi^{(0)} 
		+ \mathcal L_H^{(1)} \nabla^2 \phi^{(1)} - \nabla^2 \mathcal L_H^{(1)} \phi^{(1)} \\ 
		& \quad 	- \mathcal L_H^{(1)} RHS^{(1)} - \mathcal L_H^{(2)} RHS^{(0)}. \end{split} \label{condT2}
}
The right hand side 
is known in terms of $RHS^{(1)}$ in \eqref{RHS1known}, $RHS^{(0)}$ in \eqref{RHS0}, 
$\mathcal L_H^{(1)}$ in \eqref{LH0LH1}, $\nabla^2$ of $AdS_3$, and the assumed $\mathcal L_H^{(2)} = 0$.

Proceeding in an identical matter as at lower orders, we impose an ansatz for $RHS^{(2)}$. It will contain terms: $(\tilde T_{++}\tilde T_{--})$ multiplying ($\phi^{(0)}, \p_Z\phi^{(0)}, \p_Z^2 \phi^{(0)}, \p_+\p_-\phi^{(0)})$, and 
$\tilde T_{++}\tilde T'_{--} \p_+\phi^{(0)}$ and $\tilde T_{++}\tilde T''_{--} \p_+^2 \phi^{(0)}$ (all symmetrized over $+ \leftrightarrow -$ exchange). Said otherwise, the ansatz for $RHS^{(2)}$ will contain 8 undetermined coefficients. Additionally, it will contain terms $b \, \phi^{(2)} + 4 Z^4 \gamma (\tilde T_{++} \p_-^2 \phi^{(1)} + \tilde T_{--} \p_+^2 \phi^{(1)})$ such that the previous orders in $RHS$ are consistently extended to include higher orders in $\phi$. 

Next, we write out the left hand side and right hand side of \eqref{condT2}\footnote{
	For this, we make use of the action of the modular Hamiltonian on a general, second order derivative ansatz for the $n$-th order in $1/N$ contribution to the $RHS$ of the bulk equation of motion: 
	\ali{
		[H_{mod},RHS^{(n)}] &= (\p_{\tilde T_{++}} RHS^{(n)}) [H_{mod},\tilde T_{++}] +  (\p_{\tilde T_{--}} RHS^{(n)}) [H_{mod},\tilde T_{--}]   \nonumber \\ 
		&+ (\p_{\tilde T_{++}'}RHS^{(n)}) \p_+ [H_{mod},\tilde T_{++}] + (\p_{\tilde T'_{--}} RHS^{(n)}) \p_- [H_{mod},\tilde T_{--}] \nonumber \\ 
		&+ (\p_{\p_-\phi} RHS^{(n)}) \p_- [H_{mod},\phi] + (\p_{\p_+\phi}RHS^{(n)}) \p_+[H_{mod},\phi] + (\p_{\p_Z\phi} RHS^{(n)}) \p_Z[H_{mod},\phi]  \nonumber\\ 
		&+ (\p_{\p_+^2 \phi} RHS^{(n)}) \p_+^2 [H_{mod},\phi] + (\p_{\p_-^2\phi} RHS^{(n)}) \p_-^2 [H_{mod},\phi] \nonumber \\ 
		&+ (\p_\phi RHS^{(n)}) \, [H_{mod},\phi] + (\p_{\p_+\p_-\phi} RHS^{(n)}) \p_+\p_-[H_{mod},\phi].   
	}   
}. 
Equating the prefactors of the different derivative terms of $\phi$ 
gives rise to a set of ten equations, which 
does not have a solution for the undetermined coefficients in the ansatz for $RHS^{(2)}$.   
We conclude that the assumption $\mathcal L_H^{(2)} = 0$ was incorrect.

We can instead allow for the most general $\mathcal L_H^{(2)}$ in $[H_{mod},\phi]$ we can think of. 
Besides it being quadratic in the stress tensor, we will impose that it contains a component in $\p_+$ and one in $\p_-$, just as $\mathcal L_H^{(1)}$ in \eqref{LH0LH1} does. A $\p_Z$ term is in principle not disallowed, but since it turns out not to be necessary for finding a solution, we do not include it in the ansatz. 
The ansatz for $\mathcal L_H^{(2)}$ will then contain 6 unknown coefficients, for the possible terms obtained from multiplying $Z^6 (\tilde T_{++}\tilde T_{--},\tilde T_{++}^2,\tilde T_{--}^2)$ with $(\p_+, \p_-)$. 
Condition \eqref{condT2} can now be reordered into an unknown left hand side equal to a known right hand side: 
\ali{
	\begin{split}	\mathcal L_H^{(0)} RHS^{(2)} - \frac{1}{2\pi i} [H_{mod},RHS^{(2)}] - &(\mathcal L_H^{(2)} \nabla^2 \phi^{(0)} - \nabla^2 \mathcal L_H^{(2)} \phi^{(0)}) + \mathcal L_H^{(2)} RHS^{(0)} \\
		= & (\mathcal L_H^{(1)} \nabla^2 \phi^{(1)} - \nabla^2 \mathcal L_H^{(1)} \phi^{(1)}) 	- \mathcal L_H^{(1)} RHS^{(1)} .  \end{split} \label{condT2bis}
}
We can write out the left hand side using the ansatze for $RHS^{(2)}$ and $\mathcal L_H^{(2)}$, 
and the right hand side using $RHS^{(1)}$ in \eqref{RHS1known}, $\mathcal L_H^{(1)}$ in \eqref{LH0LH1} and $\nabla^2$ of $AdS_3$. Subsequently equating the prefactors of the different derivative terms of $\phi$ results in a set of equations 
that can uniquely 
be solved for the coefficients in the ansatze. The resulting $RHS^{(2)}$ and $\mathcal L_H^{(2)}$ are given by 
\ali{
		RHS^{(2)} &= \gamma^2  \left(  \tilde T_{--}\tilde T_{++} (4 Z^5 \p_Z\phi^{(0)} - 12 Z^6 \p_+\p_-\phi^{(0)}) -4 Z^6 (\tilde T_{--}\tilde T'_{++} \p_-\phi^{(0)} + \tilde T_{++}\tilde T'_{--} \p_+\phi^{(0)}) \right) \nonumber \\
		&\quad  + 4 Z^4 \gamma \,  (\tilde T_{++} \p_-^2 \phi^{(1)} + \tilde T_{--} \p_+^2 \phi^{(1)})  + b \, \phi^{(2)} 
}
and   
\ali{
	\mathcal L_H^{(2)}= \frac{\gamma^2}{2R}   Z^6 \tilde T_{--} \tilde T_{++} ( \p_+ - \p_-) . 
}

We have shown that the compatibility condition \eqref{Hmodconditiongrav} 
fixes \emph{both} the action of $H_{mod}$ on $\phi$ and the bulk equation of motion for $\phi$ at second order in $1/N$.

\subsection{Higher order} 

The process of fixing both the action of $H_{mod}$ on $\phi$ and the bulk equation of motion for $\phi$ at second order in $1/N$ can be repeated at each order. 
Indeed, the compatibility condition \eqref{conditiongrav} takes the form of an iterative equation for solving the $n$-th order in $\tilde T_{\pm\pm}$ contribution to the equation of motion (namely $RHS^{(n)}$) 
as well as the $n$-th order in $\tilde T_{\pm\pm}$ contribution to $[H_{mod},\phi]$ (namely $\mathcal L_H^{(n)}$), if the previous orders $RHS^{(n-1)}$, ..., $RHS^{(0)}$ and $[H_{mod},\phi]$ to $(n-1)$-st order in $\tilde T_{\pm\pm}$ 
are known: 
\ali{
	\begin{split} 
		&[\nabla^2, \mathcal L_H^{(n)}] \phi^{(0)} + \mathcal L_H^{(n)} RHS^{(0)} + \mathcal L_H^{(0)} RHS^{(n)} - \frac{1}{2\pi i} [H_{mod},RHS^{(n)}] \\ 
		&= \sum_{i=1}^{n-1} \left( [\mathcal L_H^{(i)},\nabla^2] \phi^{(n-i)} - \mathcal L_H^{(i)} RHS^{(n-i)} \right).    
	\end{split}  \label{Hmodconditionbis}
}
The right hand side of this condition is known from the previous orders, and the left hand side has unknowns $\mathcal L_H^{(n)}$ and $RHS^{(n)}$ for which we use appropriate ansatze. The ansatze we use are the most general ones, up to including at most second order 
derivatives on $\phi$ in $RHS^{(n)}$ and not including a $\p_Z$ term in $\mathcal L_H^{(n)}$.  This means that in principle there could be more consistent pairs of $H_{mod}$ action and possible equations of motion.

Let us give one more example. For $n=3$, the ansatz for the third order contribution $RHS^{(3)}$ to the equation of motion for $\phi$ will contain the expression for $RHS^{(2)}$ with $\phi^{(i)}$ replaced by $\phi^{(i+1)}$. It will also contain $\p_-\phi$ times $\tilde T_{++}^2 \tilde T_{--}'$ and $\p_-^2 \phi$ times $\tilde T_{++}^2 \tilde T_{--}$ (symmetrized over $+ \leftrightarrow -$ exchange). This leaves 4 coefficients to be determined. The ansatz for $\mathcal L_H^{(3)}$ contains 8 undetermined coefficients, for  
the possible combinations of factors $Z^8 (\tilde T_{--}^3, \tilde T_{--}^2 \tilde T_{++}, \tilde T_{--}\tilde T_{++}^2, \tilde T_{++}^3)$ and $(\p_+\phi, \p_-\phi)$.  
The condition \eqref{Hmodconditiongrav}, rewritten into \eqref{Hmodconditionbis}, uniquely determines the coefficients. The resulting addition to the equation of motion of $\phi$ is 
\ali{
	RHS^{(3)} = \gamma^3 \left( 4 Z^8 \tilde T_{++}^2 \tilde T'_{--} \p_-\phi^{(0)} + 8 Z^8 \tilde T_{++}^2 \tilde T_{--} \p_-^2 \phi^{(0)} + \,\,\, (+\leftrightarrow-) \,\,\, \right) + \cdots 
}
where $\cdots$ are the terms in $\phi^{(1)}$, $\phi^{(2)}$ and $\phi^{(3)}$ that are immediate from the previous orders in the equation of motion. 
And the resulting addition to the modular Hamiltonian action on $\phi$ is 
\ali{
	\mathcal L_H^{(3)} = \frac{\gamma^3}{2R} Z^8 (\tilde T_{++} \tilde T_{--}^2 \p_+\phi - \tilde T_{--} \tilde T_{++}^2 \p_-\phi). 
}

We now fix our normalization to $\frac{\gamma}{N^2} = \frac{6}{c}$, such that $\frac{\gamma}{N} \tilde T_{\pm\pm} = \frac{6}{c} T_{\pm\pm}$. 
The results we find at each order are consistent with the \emph{all-order}  bulk equation of motion, as well as the \emph{all-order} modular Hamiltonian action \eqref{HmodactiongravApp} that is obtained from a bulk perspective in Appendix \ref{AppBulkgravity}:    
\ali{
	\frac{1}{2\pi i} [H_{mod},\phi] = \xi^M \p_M \phi + \epsilon^+  
	\p_+\phi + \epsilon^- \p_- \phi   \label{Hmodactiongrav}
}
in terms of the stress tensor dependent functions 
\ali{
	\epsilon^+  
	= \frac{1}{2R} \frac{ \frac{6}{c} Z^4 T_{--} + \frac{36}{c^2} Z^6 T_{--}T_{++}}{1- \frac{36}{c^2}Z^4 T_{--}T_{++}}, \qquad 
	\epsilon^- = -\frac{1}{2R} \frac{\frac{6}{c} Z^4 T_{++} + \frac{36}{c^2} Z^6 T_{--}T_{++}}{1-\frac{36}{c^2} Z^4 T_{--}T_{++}}, \label{lambdafunctions} 
}
and 
\ali{
	\nabla^2 \phi - m_0^2 \, \phi &=  
	\frac{4 Z^8  \frac{6^3}{c^3} T_{++}^2 T_{--}' - 4 Z^6  \frac{36}{c^2} T_{--}T_{++}'}{(1 - Z^4  \frac{36}{c^2} T_{--}T_{++} )^3} \p_-\phi + \frac{4 Z^4  \frac{6}{c}T_{++}}{(1 - Z^4  \frac{36}{c^2}T_{--}T_{++})^2} \p_-^2 \phi \nonumber \\ 
	&+ \frac{4 Z^8  \frac{6^3}{c^3}T_{--}^2 T'_{++} - 4Z^6  \frac{36}{c^2}T_{++}T'_{--}}{(1 - Z^4  \frac{36}{c^2}T_{--}T_{++})^3} \p_+\phi + \frac{4 Z^4  \frac{6}{c}T_{--}}{(1 - Z^4  \frac{36}{c^2}T_{--}T_{++})^2} \p_+^2 \phi \nonumber \\ 
	&+ \frac{4Z^6  \frac{36}{c^2}T_{--}T_{++} (-3 + Z^4  \frac{36}{c^2}T_{--}T_{++})}{(1 - Z^4  \frac{36}{c^2}T_{--}T_{++})^2} \p_+\p_-\phi + \frac{4 Z^5  \frac{36}{c^2}T_{--}T_{++}}{1 - Z^4  \frac{36}{c^2}T_{--}T_{++}} \p_Z\phi . \label{EOMgravity} 
} 
Here we have 
$b=m_0^2$ in $RHS^{(0)}$ for a massive field, and the  
right hand side contains the $(T_{\pm\pm} \phi)$-interaction terms  
that, from the CFT point of view,  are allowed by the procedure of demanding compatibility with the modular Hamiltonian action on the dressed $\phi$. 

The all-order bulk equation of motion above is a rewriting of the, perhaps not unexpected, equation of motion of a field $\phi$ in a Banados geometry \eqref{Banados} 
\ali{
	\nabla^2_{Banados} \phi - m_0^2 \phi &= 0  
}
when the Brown-York stress tensor is replaced by the CFT stress tensor.  
We note that the full vector field $\xi + \epsilon$ in the all-order $H_{mod}$ action \eqref{Hmodactiongrav} is an asymptotic, or `Brown-Henneaux', Killing vector of the Banados geometry in the sense that it preserves holographic gauge, but is not an actual Killing vector of the Banados geometry. This is further detailed in Appendix \ref{AppBulkgravity} on the bulk perspective.

The all-order expressions \eqref{Hmodactiongrav} and \eqref{EOMgravity} for the modular Hamiltonian action and bulk equation of motion of a gravitationally dressed scalar field $\phi$ are checked to satisfy the compatibility condition \eqref{Hmodconditiongrav}.

\section{Comments}

One may think that the constraints put on the equation of motion of the bulk operator in the CFT vacuum state are a result of the modular Hamiltonian being a sum of conformal generators which are symmetries of the CFT.
We believe this is not the case, but rather that there is a more general story. Consider some non-vacuum state in the CFT which has a holographic dual. The modular Hamiltonian need not be (and is generally not) a sum of conformal generators. Nevertheless, near the HRT \cite{Hubeny:2007xt}  surface one expects the modular Hamiltonian to acts as boosts \cite{JLMS}. So very near the HRT surface it seems reasonable to expect 
\begin{equation}
	\frac{1}{2\pi i} [H_{mod},\phi]=\xi^{M} \partial_{M} \phi + \cdots
	\label{hmodgen}
\end{equation}
where $(\partial_{N}\xi^{M})|_{HRT}=\Lambda_{N}^{M}$ is the appropriate  boost matrix in the normal directions to the HRT surface. As argued in \cite{Kabat:2018smf} this constrains the background metric expected to be dual to the CFT state. In this paper, we saw that it also constrains the possible equation of motion for a bulk scalar operator, since equation (\ref{hmodgen}) has to be compatible with the equation of motion. In particular the Laplacian operator would have to be the appropriate one compatible with the metric (as we saw in equation \eqref{coneqb}  in the simple case of the vacuum state of the CFT). 
Note that this only clearly constrains the equation of motion for the state for which $H_{mod}$ was computed, since only then its action on $\phi$ is expected to have  the simple geometric form (\ref{hmodgen}) near the HRT surface. This state is of course invariant under the action of $H_{mod}$ and so one can consider this a symmetry of the particular state, but not of the whole theory. So in essence the invariance of a state under modular flow constrains the possible equation of motion of bulk operators in the dual holographic picture. What we showed in this paper is that also both general covariance and gauge invariance (in a fixed gauge) are implemented through the action of $H_{mod}$, through extra terms on the right of (\ref{hmodgen}), and our argument indicates that this should be true also in backgrounds which are not empty AdS.

In the picture of \cite{Kabat:2018smf}, the bulk spacetime manifold arises from properties of the algebra of modular Hamiltonians and the background metric arises from the compatibility of the metric $g_{MN}$ with the various boost matrices near the HRT surface (as implemented on bulk operators), representing the invariance of the state under modular flow. So the construct of a metric seems rather ``classical". In that picture a question arises: why should we consider the metric as having quantum fluctuations and why should we consider  the properly smeared energy momentum operator as this   fluctuation (in some semiclassical approximation)?  

Here we see a possible answer. The action of the modular Hamiltonian on the bulk scalar field operator interacting with gravity includes a ``classical term" which is $\xi^{M}(g_{MN}) \partial_{M} \phi$ and a quantum term $\epsilon^{M}(T_{\mu \nu})\partial_{M} \phi$.
So this structure suggests combining the action of the modular Hamiltonian as $(\xi^{M}(g_{MN})+\epsilon^{M}(T_{\mu \nu}))\partial_{M} \phi$, which looks as if smearing of $T_{\mu \nu}$ (through $\epsilon^{M}(T_{\mu \nu})$)  can be interpreted as a fluctuation of  the classical metric.
This however is not quite the case,  
rather it 
comes about in a somewhat 
indirect way through the equation of motion. 
As we saw in the $AdS_3$ case in equation \eqref{Hmodactiongrav}, 
given  a state with some expectation value  for $\langle T_{\mu \nu} \rangle$ one can combine $\xi^{M}(g_{MN})+\epsilon^{M}(\langle T_{\mu \nu} \rangle)$ but this will not be $\xi^{M}(g^{(new)}_{MN}) $. This is reasonable since the $H_{mod}$ we are dealing with is that of the vacuum state with no expectation value. What actually happens is that  the action of the modular Hamiltonian (classical plus quantum part) constrains the equation of motion, and only in the equation of motion,  putting expectation values for $T_{\mu \nu}$ translates straightforwardly into the correct  equation of motion for a scalar field with a deformed metric.


\bigskip
\goodbreak
\centerline{\bf Acknowledgements} 
\noindent
GL would like to thank D. Kabat for discussions. GL is supported in part by the Israel Science Foundation under grant 447/17. NC was supported in part by the Israel Science Foundation under grant 447/17.

\appendix  

\section{CFT calculation of the $H_{mod}$ action on $\phi^{(1)}$ in conserved current case} 
\label{AppA} 

A free bulk scalar has a CFT representation  
\ali{
	\phi^{(0)}(Z,x^+,x^-) &=  \frac{1}{2\pi} \int_{t'^2 + y'^2 < Z^2} dt' dy' \, \left( \frac{Z^2 - t'^2 - y'^2}{Z}\right)^{\Delta - 2} \mathcal O(x^+ + t'+i y',x^- - t'+i y')  \label{freephi0}
}
with normalization such that $\phi^{(0)} \stackrel{Z \ra 0}{\sim} \frac{Z^\Delta}{2(\Delta-1)} \mathcal O(x)$, and  $x^\pm = x \pm t$.

In \cite{Kabat:2020nvj}, the first order in $1/N$ correction to the CFT representation of a scalar field in $AdS_3$ interacting with a Chern-Simons 
gauge field, 
was determined to be given by a linear combination of operators ${\cal A}_{(0,l)}$  
\ali{
	\phi^{(1)} &= \sum_{l=0}^{\infty} a_l \, {\cal A}_{(0,l)}. \label{phi1sum} 	
}
The ${\cal A}_{(0,l)}$ operators are obtained from smearing the boundary operator $\p_+^{1+l} J_{(0,l)}$ as if it is a primary scalar (even though it is not actually primary)  
\ali{
	{\cal A}_{(0,l)}(Z,x^+,x^-) &=  \int_{t'^2 + y'^2 < Z^2} dt' dy' \, K_{\Delta + 2 + 2 l}
	\,  \p_+^{1+l} J_{(0,l)}(x^+ + t'+i y',x^- - t'+i y')   \label{phi1A}
}
with smearing function  
\ali{
	K_{\Delta + 2 + 2 l}(Z,x^+,x^-; t',y') = \frac{1}{2\pi} \left( \frac{Z^2 - t'^2 - y'^2}{Z}\right)^{\Delta + 2l} ,  \label{KforAl} 
}
and $J_{(0,l)}$ defined in (\ref{j0l}) as a primary, spin $(l+1)$ operator with conformal weights $(l+\frac{\Delta}{2}+1,\frac{\Delta}{2})$ made  of one conserved boundary current $j_-$, derivatives and the primary  scalar operator $\mathcal O$. 

From the standard action of the conformal generators $L_1$ 
and $\bar L_1$ 
on the current $j_-$ (with conformal weights $(1,0)$) 
and $\mathcal O$ (with conformal weights $(\frac{\Delta}{2},\frac{\Delta}{2})$), it follows that the scalar descendants $\p_+^{1+l} J_{(0,l)}$ 
transform as 
\ali{
	\bar L_1 \, \p_+^{l+1} J_{(0,l)} = \text{(primary scalar)} + (1+l)(\Delta + l) \p_+^{l} J_{(0,l)}  \label{L1onJ}
}
with (primary scalar) the transformation of $\p_+^{l+1} J_{(0,l)}$ as if it was a primary scalar, given by $(x^+)^2 \p_+ (\p_+^{l+1} J_{(0,l)}) + 2 (l+\frac{\Delta}{2}+1) x^+ (\p_+^{1+l} J_{(0,l)})$. Note that $L_1 \, \p_+^{l+1} J_{(0,l)}$ will only consist of such a (primary scalar) contribution, and so will the action of the other global conformal generators.

We set out to determine $[H_{mod},\phi^{(1)}]$ by calculating the correlator 
$\langle [H_{mod},\phi^{(1)}]\, \bar{\mathcal O} \, j_- \rangle$.  
With $H_{mod}$ defined in terms of the conformal generators in \eqref{Hmoddef} 
and knowledge of the $\bar L_1$ action \eqref{L1onJ}, we find 
\ali{
	\frac{1}{2\pi i}[H_{mod}, \p_+^{1+l} J_{(0,l)}] = 
	\frac{1}{2\pi i}[H_{mod}, \p_+^{1+l} J_{(0,l)}]_{\text{(primary scalar)}} + \frac{1}{2R} (1+l)(\Delta+l) \p_+^l J_{(0,l)}  \label{fullHmodJ}
} 
with $\frac{1}{2\pi i}[H_{mod}, \p_+^{1+l} J_{(0,l)}(t,x)]_{\text{(primary scalar)}} = \frac{1}{2R} ((t^2 + x^2 - R^2)\p_t + 2 t  x  \p_{x} + 2 t (2l+\Delta+2)) \p_+^{1+l} J_{(0,l)}$. 
It is the second contribution that carries information on the non-locality of the bulk scalar field, dressed by its gauge field interaction. We focus in the following on that non-(primary scalar) contribution,   
and want to evaluate 
\ali{
	\frac{1}{2\pi i}\left \langle [H_{mod}, \mathcal A_{(0,l)}]_{\text{non-(prim scalar)}} \, \bar{\mathcal O} \, j_- \right \rangle = \frac{(1+l)(\Delta+l)}{2R} \left \langle (\int K_{\Delta + 2 + 2 l} \, \p_+^{l} J_{(0,l)} ) \, \bar{\mathcal O} \, j_- \right \rangle  \label{tocalc}
}
at the suppressed locations $\p_+^l J_{(0,l)}(x_1^+,x_1^-)$, $\mathcal O(x_2^+, x_2^-)$ and $j_-(x_3^-)$. 
For this, we start from the CFT correlator 
\ali{
	\left \langle J_{(0,l)}(x_1^+,x_1^-) \, \bar{\mathcal O} (x_2^+, x_2^-) \, j_-(x_3^-) \right \rangle  = \frac{\alpha_l \, (-1)^{\Delta} (x_{23}^-)^l}{(x_{12}^-)^{\Delta+l} (x_{12}^+)^\Delta (x_{13}^-)^{l+2}} \label{A7}
}
with OPE coefficient $\alpha_l$, and smear it to the correlator on the right hand side of \eqref{tocalc}. We then evaluate the smearing integral in \eqref{tocalc} by a completely analogous calculation as the one detailed in Appendix A.2 of \cite{Kabat:2020nvj} for calculating $\langle \mathcal A_{(0,l)} \, \mathcal O \, j_- \rangle$. The result is 
\ali{
	&\frac{1}{2\pi i}\left \langle [H_{mod}, \mathcal A_{(0,l)}]_{\text{non-(prim scalar)}} \, \bar{\mathcal O} \, j_- \right \rangle = \frac{(1+l)(\Delta+l)}{2R} \frac{(-1)^{\Delta}\alpha_l \, (x_{23}^-)^l}{2 (x_{12}^- x_{12}^+)^{l+\Delta} (x_{13}^-)^{l+2}} \frac{\Gamma(l+\Delta) Z^{2l+\Delta+2}}{\Gamma(\Delta)(2l+\Delta+1)} \, \mathcal B,  \\
	& \mathcal B = \left( 1 + \frac{Z^2}{x_{12}^+ x_{13}^-} \right)^{-(l+\Delta)} {_2F_1}\left(l+\Delta,l+\Delta,2+2l+\Delta, \frac{\frac{-Z^2}{x_{12}^+x_{12}^-}+\frac{Z^2}{x_{12}^+x_{13}^-}}{1+\frac{Z^2}{x_{12}^+x_{13}^-}} \right).  
}
Now moving on to $\left \langle [H_{mod}, \phi^{(1)}]_{\text{non-(prim scalar)}} \, \bar{\mathcal O} \, j_- \right \rangle = \sum a_l \, \left \langle [H_{mod}, \mathcal A_{(0,l)}]_{\text{non-(prim scalar)}} \, \bar{\mathcal O} \, j_- \right \rangle$, an expression is obtained that depends on the combination $a_l \alpha_l$ of the coefficients $a_l$ in \eqref{phi1sum} and the OPE coefficients $\alpha_l$. 
The same combination appears in the correlator $\langle \phi^{(1)} \, \bar{\mathcal O} \, j_- \rangle$, when calculated similarly from $\langle \mathcal A_{(0,l)} \, \bar{\mathcal O} \, j_- \rangle$, 
as referred to in the previous paragraph. Requiring the absence of unwanted branch cuts in $\langle \phi^{(1)} \, \bar{\mathcal O} \, j_- \rangle$ (that appear because of the hypergeometric function in its expression), fixes the combination $a_l \alpha_l$ up to an $l$-independent constant $c_1$. 
This constant is determined by requiring 
the absence of an unphysical pole 
in $\langle \phi^{(0)} \, \bar{\mathcal O} \, j_- \rangle$ (see \cite{Kabat:2020nvj})  
\begin{equation}
c_1= \frac{1}{2\pi}\frac{-iq}{2\Gamma(\Delta)(\Delta-1)}.
\end{equation}
Using the expression thus obtained for $a_l \alpha_l$ in \cite{Kabat:2020nvj},   
\ali{
	a_l \alpha_l = 2 \, c_1 (1 + 2l + \Delta) \frac{\Gamma(\Delta)\Gamma(l+\Delta)}{\Gamma(1+2l+\Delta)},  
}
we arrive at 
\ali{
	\frac{1}{2\pi i}\left \langle [H_{mod}, \phi^{(1)}]_{\text{non-(prim scalar)}} \, \bar{\mathcal O} \, j_- \right \rangle &= \frac{c_1}{2R} \left(\frac{x_{13}^-}{Z}\right)^{\Delta-2} \frac{(-1)^{\Delta}}{(x_{23}^-)^\Delta} \sum_{l=0}^\infty \left\{  (l+1) \frac{\Gamma(l+\Delta+1) \Gamma(l+\Delta)}{\Gamma(1+2l+\Delta)} \right. \nonumber \\ 
	&\quad \left. \phantom{\frac{1}{2}}  \times  (-\mathcal Y)^{l+\Delta} \,  {_2F_1}\left(l+\Delta,l+2,2+2l+\Delta,\mathcal Y \right)  \right\}  
	\label{Hmodphi1correl} 
}
for 
\ali{
	\mathcal Y = Z \, \frac{x_{23}^-}{x_{13}^-} \,  \frac{Z}{Z^2 + x_{12}^- x_{12}^+}. \label{mathcalY} 
}
The sum over $l$ in \eqref{Hmodphi1correl} evaluates to $\Gamma(\Delta) \mathcal Y^\Delta$, thanks to the hypergeometric identity \cite{Hansen}  
\ali{
	\sum_k \frac{(a)_k (b)_k (c)_k}{\Gamma(k+1) (c)_{2k}} \, x^k \, F_{2,1}(a+k,b+k,c+2k+1,-x) = 1 \label{Hansensum}
}
with Pochhammer symbols $(a)_k = \frac{\Gamma(a+k)}{\Gamma(a)}$. 
The resulting correlator 
is given by 
\ali{
	\frac{1}{2\pi i}\left \langle [H_{mod}, \phi^{(1)}]_{\text{non-(prim scalar)}} \, \bar{\mathcal O} \, j \right \rangle &= \frac{c_1 \Gamma(\Delta)}{2R} \frac{Z^2}{(x_{13}^-)^2} \left( \frac{Z}{x_{12}^- x_{12}^+ + Z^2} \right)^\Delta.   \label{Hmodcorr}
}
On the other hand, we know 
\ali{
	\left \langle j_-(x_1^-) \phi^{(0)}(Z,x_1^+,x_1^-) \bar{\mathcal O} (x_2^+,x_2^-) j_-(x_3^-) \right \rangle = \frac{1}{2\pi}\frac{1}{2(\Delta-1)} \frac{k}{(x_{13}^-)^2} \left( \frac{Z}{x_{12}^- x_{12}^+ + Z^2} \right)^\Delta   \label{4ptf}
}
from contracting 
$\langle \phi^{(0)}(Z,x_1) \bar{\mathcal O}(x_2) \rangle = \frac{1}{2(\Delta-1)} \left( \frac{Z}{x_{12}^- x_{12}^+ +Z^2} \right)^\Delta$ and $\langle j_-(x_1^-) j_-(x_3^-) \rangle = \frac{k}{2\pi (x_{13}^-)^2}$.    
It follows from comparison between \eqref{Hmodcorr} and \eqref{4ptf} that 
\ali{
	\frac{1}{2\pi i}[H_{mod}, \phi^{(1)}]_{\text{non-(prim scalar)}} = - \frac{iq}{k} \frac{Z^2}{2R} \, j_- \phi^{(0)}.  
}

Now let us consider the (primary scalar) contribution to $[H_{mod}, \p_+^{1+l} J_l]$ in \eqref{fullHmodJ}. We know from  \cite{Kabat:2018smf},  that this will give rise to a corresponding (primary scalar) contribution to $[H_{mod}, \mathcal A_{(0,l)}]$ given by 
\ali{
	\frac{1}{2\pi i}[H_{mod}, \mathcal A_{(0,l)}(Z,t,x)]_{\text{(primary scalar)}} &= \xi^M \p_M \mathcal A_{(0,l)}(Z,t,x), 
}
with $\xi$ the bulk Killing vector 
\ali{
	\xi = \xi^M \p_M = \frac{1}{2R} (Z^2 + x^2 - R^2 + t^2)\p_t + \frac{t}{R} (Z\p_Z + x \p_x)  \label{bulkKilling}
}
of empty $AdS_3$ 
\ali{ 
	ds^2 = \frac{1}{Z^2} (dZ^2 - dt^2 + dx^2), \label{AdS3}
} 
that vanishes on the Ryu-Takayanagi surface \cite{Ryu:2006bv}. 
We conclude that the action of $H_{mod}$ on $\phi = \phi^{(0)} + \frac{1}{N} \phi^{(1)}$ is given by 
\ali{
	\frac{1}{2\pi i} [H_{mod},(\phi^{(0)}+\frac{1}{N}\phi^{(1)})] = \xi^M \p_M (\phi^{(0)}+\frac{1}{N}\phi^{(1)})+ \frac{1}{N} \frac{\beta Z^2}{2R} j_- \phi^{(0)}, \qquad \beta = - \frac{i q}{k}.  
}
This is equation \eqref{hmodact1}. 

\section{$H_{mod}$ action on $\phi^{(1)}$ for general spin} \label{AppB} 

We generalize here the discussion in Appendix \ref{AppA} to the case of a conserved, chiral spin $s$ current $j_{(s,0)}$ in the CFT, with $s>1$. This more general set-up was  
considered in \cite{Kabat:2020nvj}, where the $s>1$ generalization of $\phi^{(1)}$ in \eqref{phi1sum}-\eqref{KforAl} is given explicitly, 
and equation \eqref{L1onJ} generalizes to 
\ali{
	\bar L_1 \, \p_+^{s+l} J^{(s)}_{(0,l)} = \text{(primary scalar)} + (s+l)(\Delta + s+l-1) \p_+^{s+l-1} J^{(s)}_{(0,l)}  \label{L1onJgens}
}
with $J^{(s)}_{(0,l)}$ now a primary spin ($s+l$) field with conformal dimensions $(l+\frac{\Delta}{2}+s, \frac{\Delta}{2})$.   

The derivation of \eqref{Hmodphi1correl} can straightforwardly be repeated for this more general case, with the result 
\ali{
	\begin{split} 
		& \frac{1}{2\pi i} \left \langle [H_{mod}, \phi^{(1)}]_{\text{non-(prim scalar)}} \mathcal O j \right \rangle = \frac{(-1)^{\Delta}}{4R \Gamma(\Delta)} \frac{(x_{12}^-)^{s-1} (x_{13}^-)^{\Delta-s-1}}{(x_{23}^-)^{s+\Delta-1} Z^{\Delta-2}} (-\mathcal Y)^{s+\Delta-1}  \\ 
		& \quad  \times \sum_{l=0}^\infty a_l \alpha_l^{(s)} \frac{(s+l)\Gamma(l+s+\Delta)}{2s+2l+\Delta-1}  (-\mathcal Y)^l \,\,{_2F_1}\left( l+s+\Delta-1, l+2s, 2s+2l+\Delta, \mathcal Y \right), 
	\end{split}  \label{tocalcgeneraln}
}
with $\mathcal Y$ defined in \eqref{mathcalY}. 
Here $\alpha_l^{(s)}$ is the now $s$-dependent OPE coefficient in \eqref{A7},  and the combination $a_l \alpha_l^{(s)}$ has to be determined from the CFT regularity condition 
in equation (78) of \cite{Kabat:2020nvj}: 
\ali{
	\sum_{l=0}^\infty a_l \alpha_l^{(s)} (-1)^l \frac{\Gamma(l+s+\Delta) \Gamma(2l+2s+\Delta-1)}{2 \Gamma(\Delta) \Gamma(l+\Delta) \Gamma(l+s)} \,\,{_2F_1}(l+\Delta+s, 1-l-s, 1+s, z ) = 0.    \label{conditionKL}
}  
We numerically check that 
\ali{
	\begin{split}
	& {_2F_1}(-1+\Delta+s, 2-s, 1+s, z) =  \frac{2 \Gamma(\Delta) \Gamma(\Delta-1)(\Delta+s-2)(2s-1)\Gamma(s)}{\Gamma(\Delta+2s-2)}  \\ 
	& \quad \times  \sum_{l=0}^\infty \left\{\frac{ (-1)^l (2s+2l+\Delta-1) \Gamma(s+\Delta+l-1) \Gamma(2s+l) \Gamma(2s+\Delta+l-2)}{2\Gamma(\Delta) \Gamma(s+\Delta-1)\Gamma(2s)\Gamma(l+2)\Gamma(l+\Delta)\Gamma(l+s+1)} \right. \\
	&\qquad \qquad  \left. \phantom{\frac{1}{2}} \times {_2F_1}\left(l+\Delta+s, 1-l-s, 1+s, z \right) \right\}  .  
	\end{split} 
} 
Writing this identity as $F_{lhs} = F_{rhs}$, working with the derivative operator $\mathcal L = -z(1-z)\p_z^2 + [(2+\Delta)z-(1+s)] \p_z + [2(\Delta-1)+s(3-\Delta)-s^2 ]$ gives $\mathcal L \, F_{lhs}= 0$ and $\mathcal L F_{rhs} = (1+l)(-2+\Delta+l+2s) F_{rhs}$, leading to  
\ali{
	\begin{split} 
	&\sum_{l=0}^\infty \left\{ (-1)^l (2s+2l+\Delta-1) \frac{\Gamma(s+\Delta+l-1) \Gamma(2s+l) \Gamma(2s+\Delta-1+l)}{2 \Gamma(\Delta) \Gamma(s+\Delta-1) \Gamma(2s) \Gamma(l+1) \Gamma(l+\Delta) \Gamma(l+s+1)} \right. \\
	&\qquad \left. \phantom{\frac{1}{2}} \times {_2F_1}(l+\Delta+s, 1-l-s, 1+s, z ) \right\}  = 0  
	\end{split} 	
}
(where we dropped an $l$-independent constant). 
It follows from comparison with \eqref{conditionKL} that 
\ali{
	a_l \alpha_l^{(s)} =  \frac{2s+2l+\Delta-1}{s+l} \frac{ c_{s,\Delta}\, \,  \Gamma(s+\Delta-1+l) \Gamma(2s+l) \Gamma(2s+\Delta-1+l)}{\Gamma(l+s+\Delta) \Gamma(s+\Delta-1) \Gamma(2s) \Gamma(l+1) \Gamma(2s+\Delta-1+2l)}
}
where $c_{s,\Delta}$ is an $l$-independent (but in general $\Delta$- and $s$-dependent) constant.  
Plugging this into \eqref{tocalcgeneraln}, the sum on the second line of \eqref{tocalcgeneraln} evaluates to 1 by the use of \eqref{Hansensum}. 
We have thus found 
\ali{
	\frac{1}{2\pi i} \left \langle [H_{mod}, \phi^{(1)}] \mathcal O j \right \rangle &= \frac{c_{s,\Delta}}{4 R \Gamma(\Delta)} Z^{s+1} \frac{(x_{12}^-)^{s-1}}{(x_{13}^-)^{2s}} \left( \frac{Z}{x_{12}^- x_{12}^+ + Z^2} \right)^{s+\Delta-1} . \label{eqB7}
}
On the other hand, 
\ali{
	\left \langle j_{(s,0)}(x_1^-) \p_{x_1^+}^{s-1} \phi^{(0)}(Z,x_1) \mathcal O(x_2) j_{(s,0)}(x_3^-) \right \rangle \sim \frac{1}{(x_{13}^-)^{2s}} \left( \frac{x_{12}^-}{Z} \right)^{s-1} \left( \frac{Z}{x_{12}^- x_{12}^++ Z^2} \right)^{\Delta + s - 1}  \label{4ptfgenerals}
}
from contracting $\langle j_{(s,0)}(x_1^-) j_{(s,0)}(x_2^-) \rangle \sim \frac{1}{(x_{12}^-)^{2s}}$ 
and $\langle \phi^{(0)}(Z,x_1) \mathcal O(x_2) \rangle \sim \left( \frac{Z}{x_{12}^- x_{12}^+ + Z^2} \right)^\Delta$, 
and using 
\ali{
	\p_{x_1^+}^{s-1} \left( \frac{Z}{x_{12}^- x_{12}^+ + Z^2} \right)^\Delta \sim \left( \frac{x_{12}^-}{Z} \right)^{s-1} \left( \frac{Z}{x_{12}^- x_{12}^+ + Z^2} \right)^{\Delta + s - 1}.  
}
From comparison between \eqref{eqB7} and \eqref{4ptfgenerals}, we conclude 
\ali{
	\frac{1}{2\pi i} [H_{mod}, \phi^{(1)}]_{\text{non-(prim scalar)}} \sim  \frac{Z^{2s}}{2R} \, j_{(s,0)} \p_+^{s-1} \phi^{(0)}.  
}
Finally, in the presence of a conserved spin-$s$ current in the CFT, the action of $H_{mod}$ on the first order in $1/N$ CFT representation $\phi = \phi^{(0)} + \frac{1}{N} \phi^{(1)}$ of a charged bulk scalar field is given by 
\ali{
	\frac{1}{2\pi i} [H_{mod},\phi] = \xi^M \p_M \phi + \frac{1}{N} \frac{\gamma \, Z^{2n}}{2R} j_{(s,0)} \p_+^{s-1} \phi \label{Hmodjs}
}  
with a constant $\gamma$ that we leave undetermined.

\section{Bulk calculation of the $H_{mod}$ action on gauge dressed $\phi$} \label{AppC} 

In this Appendix, we use the bulk perspective to derive the $H_{mod}$ action on a charged scalar field $\phi$ in $AdS_3$ spacetime \eqref{AdS3}, with coordinates $x^M = (Z, x^+, x^-)$, that is coupled to a $U(1)$ Chern-Simons gauge field. 
The theory under consideration has the action   
\ali{
	S = \int 
	\sqrt{|g|} \left( (D_M \phi) (D^M \phi)^* + m^2 |\phi|^2 - \frac{k}{2} \epsilon^{MNP} A_M \p_N A_P \right)    
}
with covariant derivative $D_M = \p_M+ i \frac{q}{N} A_M$ (the charge being defined by equation \eqref{chargecond}). 
The field $\phi$ satisfies the bulk equation of motion 
\ali{
	(\nabla^2 - m^2) \phi = -i \frac{q}{N} \left[ (\nabla_M A^M) \phi + 2 A_M \p^M \phi \right] - \left( \frac{iq}{N} \right)^2 A_M A^M \phi   \label{EOMgauge}
}
with $\nabla^2 \phi = \frac{1}{\sqrt{-g}} \p_M \left(\sqrt{-g} \p^M \phi \right)$ and $\nabla_M A^M = \frac{1}{\sqrt{-g}} \p_M \left(\sqrt{-g} A^M \right)$.

In the bulk, there is a gauge redundancy under $A_M \ra A_M + \p_M \lambda$ and $\phi \ra e^{-i \frac{q}{N} \lambda} \phi$. A physical charged scalar field should therefore be `dressed' into a gauge invariant object. One simple choice of dressing is to attach a Wilson line that extends to the boundary along the $Z$-direction \cite{Donnelly:2015hta}
\ali{
	\Phi = e^{i \frac{q}{N} \int_0^Z A_Z(Z') dZ'} \phi.  
}
The obtained physical field $\Phi$ equals the scalar $\phi$ in the `holographic gauge' $A_Z=0$. We will choose to deal with the gauge redundancy by fixing the gauge in this way. 
Once we have done that, the field $\phi$ is physical and has a dual CFT representation. 

The holographic gauge is not preserved under all $AdS_3$ isometries. The generators of AdS isometries are given by 
\ali{
	\xi_1 &= Z x^- \p_Z + (x^-)^2 \p_- - Z^2 \p_+, \qquad \xi_0 = \frac{1}{2} Z \p_Z+ x^- \p_-, \qquad \xi_{-1} = \p_- \label{appendixeqa} \\    
	\xi_{\bar 1} &= Z x^+ \p_Z + (x^+)^2 \p_+ - Z^2 \p_-, \qquad \xi_{\bar 0} = \frac{1}{2} Z \p_Z + x^+ \p_+, \qquad \xi_{{- \bar 1}} = \p_+ .  
	\label{appendixeq} 
	}
That is, the Lie derivative of the $AdS_3$ metric \eqref{AdS3} along these vectors vanishes, $\mathcal L_{\xi_{0,\pm 1}} g_{MN} =  \mathcal L_{\xi_{\bar 0,\pm \bar 1}} g_{MN} = 0$. This is dual to the conformal invariance of the dual CFT vacuum state.   
The action of the conformal generators on a CFT representation of a free bulk scalar field 
\eqref{freephi0} accordingly is given by (see e.g.~\cite{Nakayama:2015mva,Kabat:2018smf}) 
\ali{
	[L_1,\phi^{(0)}] &= \mathcal L_{\xi_1} \phi^{(0)}, \quad [L_0,\phi^{(0)}] = \mathcal L_{\xi_0} \phi^{(0)} ,   \quad [L_{-1},\phi^{(0)}] = \mathcal L_{\xi_{-1}} \phi^{(0)} \label{C6} \\
		[\bar L_1,\phi^{(0)}] &= \mathcal L_{\xi_{\bar 1}} \phi^{(0)}, \quad [\bar L_0,\phi^{(0)}] = \mathcal L_{\xi_{\bar 0}} \phi^{(0)} , \quad [\bar L_{-1},\phi^{(0)}] = \mathcal L_{\xi_{-\bar {1}}} \phi^{(0)} .  \label{C7} 
}
The Lie derivative of the gauge field is given, in general notation, by $\mathcal L_\chi A_M = \chi^N \p_N A_M + A_N \p_M \chi^N$. It follows that the special conformal transformations $L_1$ and $\bar L_1$ (in dual conformal language) take you out of the holographic gauge 
\ali{
	\left. \mathcal L_{\xi_1} A_Z \right|_{A_Z=0} = - 2 Z A_+, \qquad \left. \mathcal L_{\xi_{\bar 1}} A_Z \right|_{A_Z=0} = - 2 Z A_-. 
	} 
Since $A_+$ and $A_-$ are canonically conjugate to each other, only one of them can fluctuate and the other is fixed as a boundary condition. We choose to set $A_+=0$. It is then only the $\bar L_1$ transformation that doesn't preserve holographic gauge.

In $A_Z=0$ gauge, the other components of the gauge field transform under AdS transformations as a combination of a Lie derivative and  a compensating gauge transformation $\lambda$, where $\lambda$ is determined by the condition \cite{KL1204,Kabat:2018smf}
\ali{ 
	\left. \mathcal L_{\xi_{\bar 1}} A_Z\right|_{A_Z=0} + \p_Z \lambda = 0.
} 

In the fixed gauge, the physical scalar field $\phi$ will then transform under AdS transformations by a combination of a Lie derivative and the compensating gauge transformation. For  instance  under the action of $\bar L_1$: 
\ali{
	[\bar L_1,\phi] = \mathcal L_{\xi_{\bar 1}} \phi - i \frac{q}{N} \bar \lambda \phi,   
	}
where 
$\p_Z \bar \lambda = 2 Z A_-$. In $AdS_3$, in particular, we have that the conserved boundary current becomes the bulk gauge field $A_\mu(Z,x^+,x^-) = \frac{1}{k} j_\mu(x^+,x^-)$, and thus 
\ali{
	\bar \lambda = \frac{1}{k} Z^2 j_-. 
}

Next, let's consider the action of $H_{mod}$, defined in \eqref{Hmoddef} in terms of the conformal generators. 
It will be given by the regular action $\mathcal L_\xi \phi$, with the bulk Killing vector $\xi$ defined in \eqref{bulkKilling}, supplemented with the above determined compensating gauge contribution 
\ali{
	\frac{1}{2\pi i}[H_{mod},\phi] = \mathcal L_\xi \phi + \frac{1}{2R N} (- i q \bar \lambda 
	) \phi 
}  
or 
\ali{
	\frac{1}{2\pi i}[H_{mod},\phi] = \xi^M \p_M \phi - \frac{i q}{k N} \frac{Z^2}{2R} j_- \phi.  
}  
This is equation \eqref{hmodact2}. 

 We can include both currents $j_-$ and $j_+$ by considering a bulk action with two Chern-Simons gauge fields $A_M$ and $\bar A_M$, with a relative sign between them, see e.g.~\cite{Jensen:2010em,Keranen:2014ava}. For the second Chern-Simons gauge field one chooses $A_-=0$, resulting in a boundary current $j_{+}$. The same reasoning as above applies and straightforwardly gives rise to 
\ali{
	\frac{1}{2\pi i} [H_{mod},\phi] = \xi^M \p_M \phi - \frac{i q}{k N} \frac{Z^2}{2R} j_- \phi + \frac{i q}{k N} \frac{Z^2}{2R} j_+ \phi \, ,   
} 
where the sign difference is due to that now it is the $L_{1}$ (and not $\bar{L}_{1}$)  transformation which gets an additional contribution.
This is just equation \eqref{hmodact2+}.

\section{Bulk calculation of the $H_{mod}$ action on gravitationally dressed $\phi$} \label{AppBulkgravity}

We use the bulk perspective to obtain the $H_{mod}$ action on a  gravitationally dressed scalar field $\phi$ in an asymptotically $AdS_3$ spacetime 
with coordinates $x^M = (Z, x^+, x^-)$. The field $\phi$ interacts with gravity, corresponding to a scalar operator $\mathcal O$ coupling to a conserved CFT stress tensor $T_\mn$ in the dual theory. The CFT perspective on this set-up is discussed in section \ref{gravsection}. 

If $\phi$ were non-interacting, the $H_{mod}$ action would be given by the Lie derivative with respect to the bulk Killing vector $\xi$ defined in \eqref{bulkKilling} \cite{JLMS,Kabat:2018smf}. Turning on gravitational interactions, there will be corrections to the $H_{mod}$ action. 
We derive in this Appendix the general, all-order form of these corrections $\Lambda$ in 
\ali{
	\frac{1}{2\pi i} [H_{mod},\phi ] = \xi^M \p_M \phi + \Lambda \, \phi.  \label{generalformApp} 
}

Diffeomorphism invariance in the bulk is invisible from the CFT perspective and the gauge should be fixed. 
The gauge choice we make is that of `holographic gauge' or Fefferman-Graham gauge, in which the asymptotically $AdS_3$ bulk metric $ds^2 = g_{MN} dx^M dx^N$ (with bulk indices $M,N = Z, +, -$ and boundary indices $\mu, \nu = +,-$) takes the form 
\ali{
	ds^2 = 	\frac{\ell_{AdS}^2}{Z^2} (dZ^2 + dx^+ dx^-) + h_\mn dx^\mu dx^\nu .  \label{FGmetric}
}
In this gauge, $h_{ZZ} = h_{Z\mu}=0$ and the remaining metric perturbations $h_\mn(Z,x^+,x^-)$ have a specific expansion in $Z$ 
\cite{deHaro:2000vlm, FG}.

The holographic gauge is not preserved under the action of the modular Hamiltonian, because it is not preserved under the action of special conformal transformations $L_1$ and $\bar L_1$.  
That is, the corresponding $AdS_3$ isometry generators $\xi_1$ and $\xi_{\bar 1}$  (as defined in \eqref{appendixeqa} and \eqref{appendixeq}) take us out of holographic gauge: 
\ali{
	\mathcal L_{\xi_1} h_{Z+} = -2 Z h_{++}, \qquad \mathcal L_{\xi_1} h_{Z-} = -2 Z h_{+-} \\
	\mathcal L_{\xi_{\bar 1}} h_{Z+} = -2 Z h_{+-}, \qquad \mathcal L_{\xi_{\bar 1}} h_{Z-} = -2 Z h_{--}, 
}
with $\mathcal L_\chi h_{MN} = \nabla_M\chi_N + \nabla_N \chi_M$ the Lie derivative along the vector $\chi$. The same will be true for the bulk Killing vector $\xi$ associated with $H_{mod}$, defined in \eqref{bulkKilling}. 

In holographic gauge, the isometry action 
\ali{ 
	\mathcal L_{\xi_1 + \epsilon_1} h_{Z\mu} = 0,  \quad \mathcal L_{\xi_{\bar 1} + \epsilon_{\bar 1}} h_{Z\mu} = 0   \label{isometry}
} 
therefore includes compensating gauge transformation contributions of the form $\mathcal L_\epsilon$, with the vectors  $\epsilon_1$ and $\epsilon_{\bar 1}$ determined by the condition \eqref{isometry} and that they vanish at the boundary $Z \ra 0$.

This determines the compensating gauge vector field $\epsilon_1$ to be of the form 
\ali{
	\epsilon_1 &= Z c_1 \p_Z +  \int_0^Z dz \, \mathcal F\left( h_\mn(z,x^+,x^-); c_1\right)  \p_+  + \int_0^Z dz \, \tilde {\mathcal F} \left( h_\mn(z,x^+,x^-); c_1 \right)  \p_- \label{eqD8}
}
with $c_1$ an arbitrary function $c_1(x^+,x^-)$, and $\mathcal F$ and $\tilde {\mathcal F}$ given explicitly by 
\ali{
	\mathcal F &= \frac{-2 z \left(h_{--} \left(4 z^4 h_{++}-2 z^2 \p_+c_1\right)+\p_-c_1\right)-4 z^3 h_{+-} \left(\p_-c_1-1\right)+8 z^5 h_{+-}^2}{-4 z^4 h_{--} h_{++}+4 z^4 h_{+-}^2+4 z^2 h_{+-}+1} \\ 
	\tilde {\mathcal F} &=\frac{4 z^3 \left(\p_-c_1+1\right) h_{++}-2 z \p_+ c_1 \left(2 z^2 h_{+-}+1\right)}{-4 z^4 h_{--} h_{++}+4 z^4 h_{+-}^2+4 z^2 h_{+-}+1}. 
}

In three bulk dimensions with a negative cosmological constant, the most general solution of vacuum Einstein equation in the Fefferman-Graham expansion, 
takes a truncated form, known as the 
Banados geometry \cite{Banados:1998gg} 
\ali{
	\hspace{-0.5cm} ds^2 = 	\frac{\ell_{AdS}^2}{Z^2} (dZ^2 + dx^+ dx^-) + \frac{6}{c} \vev{T_{++}} (dx^+)^2 + \frac{6}{c} \vev{T_{--}} (dx^-)^2 + \frac{36}{c^2} \vev{T_{++}} \vev{T_{--}} Z^2 dx^+ dx^- . \label{Banados}
}
Here we have already written the 3-dimensional Newton constant in terms of the CFT central charge using $c = \frac{3 \ell_{AdS}}{2G}$, and $\vev{T_{\pm\pm}}$ are the Brown-York stress tensor components of asymptotically $AdS_3$ gravity.  
The Banados metric gives the most general solution of vacuum Einstein equations with negative cosmological constant. It captures the full non-linear interactions of the metric perturbations $h_\mn$, but does not take into account backreaction of $\phi$ on the metric. 
The field $\phi$ then gives rise to well-defined CFT correlators with any number of CFT stress tensors $\langle \phi \, \mathcal O \, T_{--} \cdots T_{++} \cdots \rangle$ (but no other operator insertions). 
Said otherwise, using the Banados metric is the gravity equivalent of approximating the gauge field $A_\mu$ by its non-backreacted (i.e.~non-corrected with $1/N$ corrections) form in section \ref{sect:allorders}.

In the same way that the bulk gauge field $A_\mu$ of section \ref{sect:allorders} asymptotes to the CFT current $j_\mu$, we have that the bulk metric perturbation $h_\mn$ asymptotes to the CFT stress tensor $T_\mn$ at leading order in $1/N$ \cite{KL1204}. 
We accordingly will interpret $T_\mn$ in the Banados geometry as a CFT operator, 
following e.g.~the interpretation in \cite{Kaplan1708}. 
The Banados metric allows us to read off the specific relation between the metric perturbations $h_\mn$ and the CFT stress tensor $T_\mn$.

Using the Banados relation between $h_\mn$ and $T_\mn$, we find $\epsilon_1$ and $\epsilon_{\bar 1}$ are given by 
\ali{
	\epsilon_1 &=  \frac{Z^4 \frac{6}{c}T_{++}}{1 - Z^4 \frac{36}{c^2} T_{--} T_{++}} \p_-  - \frac{Z^6 \frac{36}{c^2} T_{--}T_{++}}{1 - Z^4 \frac{36}{c^2}T_{--}T_{++}} \p_+\\ \label{eps1gravity} 
	\epsilon_{\bar 1} &=  \frac{Z^4 \frac{6}{c}T_{--}}{1 - Z^4 \frac{36}{c^2} T_{--} T_{++}} \p_+ - \frac{Z^6 \frac{36}{c^2} T_{--}T_{++}}{1 - Z^4 \frac{36}{c^2}T_{--}T_{++}} \p_- \, . 
}
The function $c_1$ in \eqref{eqD8} is fixed to zero by imposing that the Banados form \eqref{Banados} of the geometry is preserved. 
The resulting `new' $\xi_1$ in holographic gauge, i.e.~respecting $\mathcal L_{\xi_1}h_{Z\mu} = 0$, is given by the combination $\xi_1 + \epsilon_1 \ra \xi_1$,   
\ali{
	\xi_1 &= Z x^- \p_Z + \left( (x^-)^2 + \frac{Z^4 \frac{6}{c}T_{++}}{1 - Z^4 \frac{36}{c^2} T_{--} T_{++}} \right) \p_- - \frac{Z^2}{1 - Z^4 \frac{36}{c^2}T_{--}T_{++}} \p_+. \label{L1gravity}
}
The expression for $\xi_{\bar 1}$ is obtained from the one for $\xi_1$ by changing all $\pm$ indices to $\mp$. Equation \eqref{L1gravity} was previously written down in \cite{Kaplan1712} and in \cite{Roberts:2012aq}, where a more general expression for 
$\xi_n$ transformations ($n$ integer) can be found, corresponding to conformal transformations generated by Virasoro generators $L_n$ 
on the boundary.

Now let us go back to the action of the modular Hamiltonian on the gravitationally interacting scalar field $\phi$. 
Under a diffeomorphism $\chi$, the scalar field $\phi$ 
transforms to  $\phi + \mathcal L_\chi \phi = \phi + \chi^M \p_M\phi$. 
It follows that the compensating gauge transformations $\epsilon_1$ and $\epsilon_{\bar 1}$ will 
give rise to a contribution $\epsilon^M \p_M \phi$    
to the modular Hamiltonian action  on $\phi$ through the $\frac{1}{2R}(\bar L_1 - L_1)$ term in \eqref{Hmoddef}: 
\ali{
	\frac{1}{2\pi i} [H_{mod},\phi ] = \xi^M \p_M \phi + \epsilon^M \p_M \phi. 
} 
Written out, 
\ali{
	\frac{1}{2\pi i} [H_{mod},\phi] = \xi^M \p_M \phi + \epsilon^+ 
	\p_+\phi + \epsilon^- \p_- \phi   \label{HmodactiongravApp}
}
with $\epsilon^+ = \frac{1}{2R} (\epsilon_{\bar 1}^+ - \epsilon_1^+)$ and $\epsilon^- = \frac{1}{2R} (\epsilon_{\bar 1}^- - \epsilon_1^-)$  
given by 
\ali{
	\epsilon^+ 
	= \frac{1}{2R} \frac{ \frac{6}{c} Z^4 T_{--} + \frac{36}{c^2} Z^6 T_{--}T_{++}}{1- \frac{36}{c^2}Z^4 T_{--}T_{++}}, \qquad 
	\epsilon^- = -\frac{1}{2R} \frac{\frac{6}{c} Z^4 T_{++} + \frac{36}{c^2} Z^6 T_{--}T_{++}}{1-\frac{36}{c^2} Z^4 T_{--}T_{++}}. \label{lambdafunctionsApp} 
}
We have found that the corrections $\Lambda \phi$ in \eqref{generalformApp} take the form of compensating gauge transformation contributions $\mathcal L_\epsilon \phi$, which express the fact  that physical bulk field operators are not local. 

The result \eqref{HmodactiongravApp} is consistent with the first order in $1/N$ expression \eqref{HmodgravCFT} and the order by order expressions for $[H_{mod},\phi]$ obtained from a CFT point of view in section \ref{gravsection}. While we have used the Banados geometries to deduce the action of $H_{mod}$ on $\phi$, one has to keep in mind that this is $H_{mod}$ of the CFT vacuum state with zero expectation values for $T_{\mu \nu}$. All the terms in the $\frac{1}{N}$ expansion involving $\epsilon^{\pm}$ are not due to corrections to the modular Hamiltonian but due to corrections to the bulk operator $\phi$.

We note also that one can of course rewrite the $H_{mod}$ action as 
\ali{
	\frac{1}{2\pi i}[H_{mod},\phi] = \xi_H^M \p_M \phi  \label{Hmodactiongrav2} 
} 
in terms of a vector $\xi_H$ that is composed of $\xi$ in \eqref{bulkKilling} and  \eqref{lambdafunctionsApp} treated as some classical functions. 
This vector is an asymptotic Killing vector of the Banados geometry \eqref{Banados}: it preserves holographic gauge per construction $\mathcal L_{\xi_H} g_{ZZ} = \mathcal L_{\xi_H} g_{Z\mu} = 0$, keeping the metric of the form \eqref{FGmetric}. 
 However, $\xi_H$ is not a Killing vector of the full Banados geometry since $\mathcal L_{\xi_H} g_{\pm\pm} \neq 0$ and $\mathcal L_{\xi_H} g_{+-} \neq 0$. In particular, 
\ali{
	\mathcal L_{H} g_{--} = \frac{6}{c} [H_{mod},T_{--}], \quad \mathcal L_{H} g_{++} = \frac{6}{c} [H_{mod},T_{++}], \quad \mathcal L_{H} g_{+-} = \frac{36}{c^2} Z^2 [H_{mod},T_{++} T_{--}]  \label{LHg}
} 
such that the Banados metric is transformed under $H_{mod}$ into a new Banados metric with  
transformed stress tensors. The action of $H_{mod}$ on the stress tensors in \eqref{LHg} follows immediately from \eqref{L1Tpp}-\eqref{L1Tmm}.

\section{ $\phi^{(1)}$ comparison to charge distribution approach \label{klkapcomp}}  

In this Appendix we show how one gets the bulk field to first non-trivial order in $\frac{1}{N}$ 
from the charge distribution condition \cite{Chen:2019hdv} 
\begin{equation}
	[j_{-}(x^{-}),(\phi^{(0)}+\frac{1}{N}\phi^{(1)})(Z,y^{-},y^{+})]=-\frac{q}{N}\delta(x^{-}-y^{-})\phi^{(0)}.
	\label{bjo}
\end{equation}
The known CFT representation for the free bulk field $\phi^{(0)}$ can be written either as a smearing integral expression \eqref{freephi0} or as an expansion in the holographic radius $Z$ \cite{Nakayama:2015mva,CarneirodaCunha:2016zmi,Kaplan1708}:  
\begin{equation}
	\phi^{(0)}=\sum_{n=0}^{\infty} \alpha_{n}  Z^{2n}(\partial_{-}\partial_{+})^{n}{\cal O}  \label{freephi0sum}
\end{equation}
with
\begin{equation}
	\alpha_{n}=\frac{\Gamma(\Delta)}{2(\Delta-1)}Z^{\Delta}(-1)^{n}\frac{1}{\Gamma(n+1)\Gamma(\Delta+n)}. 
\end{equation}
Further, we know the CFT commutation relations
\begin{equation}
	[j_{-}(x^{-}), {\cal O}(y^{-},y^{+})]=-\frac{q}{N}\delta(x^{-}-y^{-}){\cal O}\ , \ \  [j_{-}(x^{-}), j_{-}(y^{-})]=ik \partial_{x^{-}} \delta(x^{-}-y^{-}).
\end{equation}

We start with
\begin{equation}
	[j_{-},\phi^{(0)}]=-\frac{q}{N}\delta(x^{-}-y^{-})\phi^{(0)} -\frac{q}{N}\sum_{n=0}^{\infty} \alpha_{n} Z^{2n} \sum_{m=1}^{n}  \frac{\Gamma(n+1)}{\Gamma(m+1)\Gamma(n-m+1)} \partial^{m}_{y^{-}} \delta(x^{-}-y^{-}) \partial^{n-m}_{-} \partial_{+}^{n} {\cal O}
	\label{jphi0}
\end{equation}
and see that we need to cancel the second term on the right hand side, which contains derivatives of delta functions of increasing order. In fact from this point of view, the purpose of the correction term in the bulk operator is to cancel unwanted parts of the charge distribution of $\phi^{(0)}$.

The most general objects that we can put into $\phi^{(1)}$ which are boundary scalars and have the right scaling dimensions under dilatations are of the form
\begin{equation}
	\phi^{(1)}=\sum_{l=0}^{\infty} Z^{2l+2+\Delta}\sum_{r=0}^{l} a_{lr}\partial_{-}^{r} j_{-} \partial_{-}^{l-r} \partial_{+}^{l+1} {\cal O}.  \label{phi1exp}
\end{equation}

Now to order $\frac{1}{N}$ the contribution of $[j_{-}, \frac{1}{N}\phi^{(1)}]$ comes from the $[j_{-}, j_{-}]$ commutator\footnote{ The $[j_{-}, {\cal O}]$ commutator is down by $\frac{1}{N}$ compared to the current-current commutator.}, thus we find to this order
\begin{equation}
	[j_{-}, \frac{1}{N}\phi^{(1)}]=-\frac{ik}{N}\sum_{l=0}^{\infty} Z^{2l+2}\sum_{r=0}^{l}  a_{lr} \partial^{r+1}_{y^{-}} \delta(x^{-}-y^{-}) \partial^{l-r}_{-} \partial_{+}^{l+1} {\cal O} + O(\frac{1}{N^2}).
\end{equation}
To get (\ref{bjo}) one needs 
\begin{equation}
	a_{lr}=i\frac{q}{k} \frac{\Gamma(\Delta)}{2(\Delta-1)}(-1)^{l+1}\frac{1}{\Gamma(\Delta+l+1)\Gamma(r+2)\Gamma(l-r+1)}.
\end{equation}
This agrees with computations in \cite{Chen:2019hdv}\footnote{
	In particular, their definition of charge $q$ maps to $-\frac{i q}{k}$ in our notation, and our normalization of $\phi^{(0)}$ in \eqref{freephi0} differs by a factor $2(\Delta-1)$ from the normalization in \cite{Chen:2019hdv}. 
}.

Note that since $ [j_{-}(x^{-}), j_{-}(y^{-})]=ik \partial_{x^{-}} \delta(x^{-}-y^{-})$ one cannot cancel the first term in equation (\ref{jphi0}), which does not have any derivative on the delta function. 
This means that one cannot construct a bulk operator $\phi$ such that $[j_{-}, \phi]=0$. This is the source of the bulk Gauss law  in this formulation.

Let us also show that the obtained double sum expression for $\phi^{(1)}$ in \eqref{phi1exp} matches the smearing integral expression for $\phi^{(1)}$ in \eqref{phi1sum}-\eqref{phi1A}, as obtained in \cite{Kabat:2020nvj}. This proceeds in the same way as showing the equivalence between expressions \eqref{freephi0} and \eqref{freephi0sum} for $\phi^{(0)}$, which involves expanding the operator $\mathcal O(x^+ + i y'+t', x^-+iy'-t')$ in the integrand around $\mathcal O(x^+,x^-)$ and subsequently working out the integral over $y'$ and $t'$ explicitly (making use of $\int_0^1 dx \, (1-x)^p x^q = \frac{\Gamma(1+q)\Gamma(1+p)}{\Gamma(2+q+p)}$). This is detailed for example in \cite{CarneirodaCunha:2016zmi}.  

The integral expression for $\phi^{(1)}$ takes the form \eqref{phi1sum}-\eqref{phi1A}, 
\ali{
	\phi^{(1)} = \sum_{l=0}^\infty  \sum_{n=0}^l a_l d_{ln} \, \int_{t'^2 + y'^2 < Z^2} dt'dy' \, \frac{1}{2\pi} \left(\frac{Z^2 - t'^2-y'^2}{Z} \right)^{\Delta+2l} \, \p_-^n j_-  \p_-^{l-n} \p_+^{l+1} \mathcal O , 
}
where the operator in the integrand is evaluated at the location $(x^++t'+i y', x^--t'+i y')$, and $d_{ln}$ is defined in \eqref{dk}. 
The coefficients $a_l$ can be determined either by solving the bulk equation of motion or by imposing the behavior under the modular Hamiltonian action given in \eqref{hmodact2}, leading to 
\ali{
	a_l &= \frac{i q}{k} \frac{\Gamma(\Delta)}{\Gamma(l+2)\Gamma(\Delta+l+1)} 
	\frac{\sum_{k=0}^l d_{lk} (2)_k (\Delta)_{l-k} (\Delta+l-k)_l}{\sum_{k=0}^l 
		\sum_{p=0}^l d_{lk} d_{lp} (2)_p (2+p)_k (\Delta)_{l-p} (\Delta+l-p)_{l-k}}    
} 
written in terms of Pochhammer symbols. 

We can first rewrite $\phi^{(1)}$ in terms of two infinite sums by applying the sum switching identity $\sum_{l=0}^\infty \sum_{n=0}^l = \sum_{n=0}^\infty \sum_{l=n}^\infty$ and then setting $l = m+n$: 
\ali{
	\phi^{(1)} = \sum_{n=0}^\infty \sum_{m=0}^\infty a_{m+n} d_{(m+n)n} \, \int_{t'^2 + y'^2 < Z^2} dt'dy' \, \frac{1}{2\pi} \left(\frac{Z^2 - t'^2-y'^2}{Z} \right)^{\Delta+2(m+n)} \, \p_-^n j_-  \p_-^{m} \p_+^{m+n+1} \mathcal O. 
}
Next, we expand the operator in the integrand around the location $(x^+,x^-)$ and explicitly evaluate the integral over $t'$ and $y'$. In the expansion, derivatives on the composite operator are worked out using $\p^i (f\, g) = \sum_{j=0}^i \frac{\Gamma(i+1)}{\Gamma(j+1)\Gamma(i-j+1)} \p^{i-j} f \p^j g$. Redefining indices $N$ and $M$ that count the powers of $\p_-$ derivatives in the composite operator, the integral expression for $\phi^{(1)}$ is rewritten as a four-sum expression: 
\ali{
	\phi^{(1)} = \sum_{N=0}^\infty \sum_{M=0}^\infty c_{NM} Z^{2N+2M+2 +\Delta} \p_-^N j_- \p_-^M \p_+^{N+M+1} \mathcal O (x^+,x^-) 
}  
with 
\ali{
	c_{NM} = \sum_{n=0}^N \sum_{m=0}^M a_{m+n} d_{(m+n)n} \frac{\Gamma(\Delta + 2n+2m+2) (-1)^{M+N+1}}{\Gamma(N+M+n+m+2+\Delta)(M-m)!(N-n)!}.  
}
It can be checked numerically that $c_{r,l-r} = a_{lr}$ so that we find agreement with expression \eqref{phi1exp} for $\phi^{(1)}$ after relabeling the indices $N \ra r$ and $M \ra l-r$. This concludes the demonstration of equivalence of the smearing integral and double sum expressions for $\phi^{(1)}$. 

We note that a straightforward generalization of the above procedure to the spin $2$ case leads to agreement between the first order gravitationally interacting field $\phi^{(1)}$ as a smearing integral \cite{Kabat:2020nvj} and as a double sum expression \cite{Kaplan1708}.

\section{Summary of used operator actions}

Scalar CFT operator $\mathcal O$ with conformal dimensions $(h_-,h_+)$ transforms as 
\ali{
	[\bar L_1,\mathcal O] &= \left((x^+)^2 \p_+ + 2 x^+ h_+ \right) \mathcal O,    
	\qquad [\bar L_0, \mathcal O] = x^+ \p_+ \mathcal O + h_+ \mathcal O, \qquad [\bar L_{-1}, \mathcal O] = \p_+ \mathcal O \\
	[L_1,\mathcal O] &= \left((x^-)^2 \p_- + 2 x^- h_- \right) \mathcal O,    
	\qquad [L_0, \mathcal O] = x^- \p_- \mathcal O + h_- \mathcal O, \qquad [L_{-1}, \mathcal O] = \p_- \mathcal O. 
}
 
\subsection{Chern-Simons gauge field} 

Scalar bulk field $\phi$ and CFT current $j$ transform as 
\ali{
	[\bar L_1,\phi] &= ( Z x^+ \p_Z + (x^+)^2 \p_+ - Z^2 \p_-)\phi + \frac{\beta}{N} Z^2 j_- \phi  \\ 
	[L_1,\phi] &= ( Z x^- \p_Z + (x^-)^2 \p_- - Z^2 \p_+)\phi +  \frac{\beta}{N} Z^2 j_+ \phi \\
	[\bar L_0,\phi] &= (\frac{1}{2}Z \p_Z + x^+ \p_+) \phi , \qquad [\bar L_{-1},\phi] = \p_+\phi \\
	[L_0,\phi] &= (\frac{1}{2}Z \p_Z + x^- \p_-) \phi, \qquad  [L_{-1},\phi] = \p_- \phi
}

\ali{
	\frac{1}{2\pi i}[H_{mod},\phi]   
	=  \xi^M \p_M \phi + \frac{\beta}{N} \frac{Z^2}{2R} \, j_- \phi - \frac{\beta}{N} \frac{Z^2}{2R} \, j_+ \phi  
}
\ali{
	C_2 \circ \phi &= \nabla^2 \phi -  \frac{2 \beta Z^2}{N} j_+ \p_-\phi  - \frac{2 \beta Z^2}{N} j_- \p_+\phi
} 
\ali{
	[\bar L_1,j_+] &= \left((x^+)^2 \p_+ + 2 x^+ \right) j_+,    
	\qquad [\bar L_0, j_+] =  x^+ \p_+ j_+ + j_+, \qquad [\bar L_{-1}, j_+] = \p_+ j_+ \\
	[L_1,j_+] &= 0,    
	\qquad [L_0, j_+] = 0, \qquad [L_{-1},j_+] = 0 \\
	[L_1,j_-] &= \left((x^-)^2 \p_- + 2 x^- \right) j_-,    
	\qquad [L_0, j_-] = x^- \p_- j_- + j_-, \qquad [L_{-1}, j_-] = \p_- j_-, \\ 
	[\bar L_1,j_-] &= 0,    
	\qquad [\bar L_0, j_-] = 0, \qquad [\bar L_{-1},j_-] = 0 
}

\subsection{Gravity} 

Scalar bulk field $\phi$ and CFT stress tensor $T_\mn$ transform as 
\ali{
	[\bar L_1,\phi] &= ( Z x^+ \p_Z + (x^+)^2 \p_+ - Z^2 \p_-)\phi + [\bar L_1,\phi]_{\text{c.g.}} \\
	[L_1,\phi] &= ( Z x^- \p_Z + (x^-)^2 \p_- - Z^2 \p_+)\phi + [L_1,\phi]_{\text{c.g.}}\\
	[\bar L_0,\phi] &= (\frac{1}{2}Z \p_Z + x^+ \p_+) \phi , \qquad [\bar L_{-1},\phi] = \p_+\phi \\
	[L_0,\phi] &= (\frac{1}{2}Z \p_Z + x^- \p_-) \phi, \qquad  [L_{-1},\phi] = \p_- \phi
}
with compensating gauge transformation contributions 
\ali{
	[\bar L_1,\phi]_{\text{c.g.}} &= \lambda_- \p_+\phi - \omega \,  \p_- \phi  \\
	[L_1,\phi]_{\text{c.g.}} &= \lambda_+ \p_- \phi - \omega \, \p_+\phi  
}
\ali{
	\lambda_- &= \frac{Z^4 T_{--}}{\mathcal N}, \qquad \lambda_+ = \frac{Z^4 T_{++}}{\mathcal N}, \qquad \omega = \frac{Z^6 T_{--} T_{++}}{\mathcal N}, \qquad \mathcal N = 1 - Z^4 T_{--}T_{++} 
} 

\ali{
	\frac{1}{2\pi i} [H_{mod},\phi] &= \xi^M \p_M \phi + \frac{1}{2R} [\bar L_1 - L_1,\phi]_\text{c.g.}     
	=  \xi^M \p_M \phi + \frac{\lambda_- + \omega}{2R} \p_+\phi - \frac{\lambda_+ + \omega}{2R} \p_- \phi    
}

\ali{
	C_2 \circ \phi 
	= 	\nabla^2 \phi -2 \p_+ \left(  \lambda_- \p_+\phi - \omega \,  \p_- \phi \right) -2 \p_- \left( \lambda_+ \p_- \phi - \omega \, \p_+\phi \right) 
}

\ali{
	[\bar L_1,T_{++}] &= \left((x^+)^2 \p_+ + 4 x^+ \right) T_{++},    
	\qquad [\bar L_0, T_{++}] =  x^+ \p_+ T_{++} + 2 T_{++}, \qquad [\bar L_{-1}, T_{++}] = \p_+ T_{++} \nonumber \\
	[L_1,T_{++}] &= 0,    \label{L1Tpp}
	\qquad [L_0, T_{++}] = 0, \qquad [L_{-1},T_{++}] = 0 \\
	[L_1,T_{--}] &= \left((x^-)^2 \p_- + 4 x^- \right) T_{--},    
	\qquad [L_0, T_{--}] = x^- \p_- T_{--} + 2T_{--}, \qquad [L_{-1}, T_{--}] = \p_- T_{--}, \nonumber \\ 
	[\bar L_1,T_{--}] &= 0,    
	\qquad [\bar L_0, T_{--}] = 0, \qquad [\bar L_{-1},T_{--}] = 0  \label{L1Tmm}
}

\bibliographystyle{JHEP}
\bibliography{referencesHmod}

\end{document}